\documentclass[11pt,a4paper]{article}

\usepackage{a4wide}
\usepackage{graphicx}
\usepackage{float}
\usepackage{algorithm}
\usepackage{algpseudocode}
\usepackage{algorithmicx}%

\usepackage{paralist}

\usepackage{tikzpagenodes}
\usetikzlibrary{fit}
\usepackage{subcaption} 
\usepackage{booktabs} 

\usepackage{amsmath}
\usepackage{url}
\usepackage{mathrsfs}
\usepackage{algorithm}
\usepackage{algorithmicx}
\usepackage{algpseudocode}
\usepackage{multirow}
\usepackage[countmax]{subfloat}
\usepackage[hidelinks]{hyperref}

\usepackage[style=numeric-comp, sorting=none]{biblatex}
\title{\textbf{Predicting the genetic component of gene expression using gene regulatory networks}}

\author{Gutama Ibrahim Mohammad$^1$ and Tom Michoel$^{1,\ast}$}

\date{}

\begin{document}

\maketitle

$^1$ Computational Biology Unit, Department of Informarics, University of Bergen, Norway

$^\ast$ Corresponding author, email: \texttt{tom.michoel@uib.no}

\begin{center}
    \textbf{Abstract}
\end{center}

\textbf{Motivation:} Gene expression prediction plays a vital role in transcriptome-wide association studies (TWAS), which seek to establish associations between tissue gene expression and complex traits. Traditional models rely on genetic variants in close genomic proximity to the gene of interest to predict the genetic component of gene expression. In this study, we propose a novel approach incorporating distal genetic variants acting through gene regulatory networks (GRNs) into gene expression prediction models, in line with the omnigenic model of complex trait inheritance.

\textbf{Results:} Using causal and coexpression GRNs reconstructed from genomic and transcriptomic data and modeling the data as a Bayesian network jointly over genetic variants and genes, inference of gene expression from observed genotypic data is achieved through a two-step process. Initially, the expression level of each gene in the network is predicted using its local genetic variants. The residuals, calculated as the differences between the observed and predicted expression levels, are then modeled using the genotype information of parent and/or grandparent nodes in the GRN. The final predicted expression level of the gene is obtained by summing the predictions from the local variants model and the residual model, effectively incorporating both local and distal genetic influences. Using various regularized regression techniques for parameter estimation, we found that GRN-based gene expression prediction outperformed the traditional local-variant approach on simulated data from the DREAM5 Systems Genetics Challenge and real data from the Geuvadis study and an eQTL mapping study in yeast. This study provides important insights into the challenge of gene expression prediction for TWAS. It reaffirms the importance of GRNs for understanding the genetic effects on gene expression and complex traits more generally.

\textbf{Availability:} The code is available on GitHub at: \href{https://github.com/guutama/GRN-TI.git}{github.com/guutama/GRN-TI}.


\newpage

\section{Introduction}

Genome-wide association studies (GWAS) have revealed that the majority of genetic variants linked to complex traits are located outside of protein-coding regions  \cite{uffelmann_genome-wide_2021}. It is widely believed that these variants primarily influence complex traits by regulating the expression of genes in a cell type specific manner \cite{albert_role_2015,pai2015genetic}, but identifying the most likely affected gene for a given GWAS variant remains a challenging task. Transcriptome-wide association studies (TWAS) aim to address this challenge by directly detecting associations between gene expression, or more precisely the genetic component of gene expression, and complex traits \cite{mai2023transcriptome}. However, transcriptomic data are rarely if ever available in the same cohort where a GWAS is performed, due to the cost of obtaining multiple omics measurements and particularly the inaccessibility of most trait-relevant tissues in living donors. 

Thus, TWAS usually involve a three-step procedure \cite{mai2023transcriptome}. First, a model to predict gene expression from genotype data is learned from smaller cohorts where the necessary transcriptomic data is available from either post-mortem samples or samples obtained during surgery. Next, the trained model is applied to predict gene expression from the genetic data available in the much larger GWAS cohorts, a process called gene expression or transcriptome imputation (TI). Finally, the predicted gene expression levels are used to assess the relationship between genes and complex traits. Because the prediction models only use genetic variants as predictors, the predicted or imputed expression levels represent the genetic component of gene expression. In this paper, we are concerned with the first step, learning models to predict the genetic component of gene expression from genomic and transcriptomic data.

To date, all prediction models used in TWAS are based on genetic variants, usually single nucleotide polymorphisms (SNPs), that are in close genomic proximity to the gene of interest and significantly associated with the gene's expression levels, so-called cis-acting expression quantitative trait loci (eQTL/cis-eQTL) \cite{albert_role_2015,pai2015genetic}. The imputation of gene expression is accomplished by utilizing individual-level genotypes of regulatory variants or summary-level data, depending on the method. 

PrediXcan \cite{gamazon_gene-based_2015} trained lasso and elastic net regression models on datasets where individual-level gene expression and SNP data are available for both the model training and subsequent TI. FUSION \cite{gusev_integrative_2016} found that a Bayesian linear mixed model trained on individual-level data using all cis-eQTLs performed best and focused on performing TI using GWAS summary statistics only. MetaXcan \cite{barbeira_exploring_2018} showed that the models developed in PrediXcan can  be fitted using cis-eQTL summary statistics and that the subsequent TI can also be carried out using GWAS summary statistics, greatly expanding the applicability of TWAS.

However, genes do not act in isolation and their expression is determined by their position in larger gene regulatory networks (GRNs) \cite{schadt2009molecular}. Reconstructing such GRNs from cohorts with both genomic and transcriptomic data using methods that can trace the chains of directionality and causality \cite{schadt2005integrative,millstein2009disentangling,chen2007harnessing,wang_efficient_2017} has led to numerous semi-mechanistic insights in the relation between GWAS variants and pathways affecting complex human diseases \cite{schadt2008, zhang2013integrated, talukdar2016, beckmann2020multiscale, koplev2022mechanistic}. Moreover, such GRNs capture a major portion of heritability beyond that of individual GWAS loci \cite{zeng2019contribution, koplev2022mechanistic}, in line with the model of omnigenic inheritance which emphasizes the importance of GRNs for understanding GWAS results \cite{boyle_expanded_2017,liu_trans_2019}. 

According to the omnigenic model, the most influential SNPs for a phenotype are enriched in specific genes or pathways that directly contribute to the phenotype (``core'' genes). On the other hand, the SNPs that account for the majority of heritability are spread across the genome, affect other genes expressed in relevant cell types (``peripheral'' genes), and exert their effect on the phenotype indirectly by affecting the expression of core genes through densely connected GRNs.

We reasoned that incorporating the principles of omnigenic inheritance in gene prediction models, that is, training gene prediction models on cis-eQTLs as well as upstream trans-eQTLs acting through GRNs, should lead to improved models for downstream TWAS compared to existing models based on cis-eQTLs alone. To test this hypothesis, we reconstructed linear Gaussian Bayesian networks using both causal and coexpression gene networks inferred from individual-level genotype and gene expression data using Findr \cite{wang_efficient_2017, wang_high-dimensional_2019}. The directed acyclic graph (DAG) structure of Bayesian networks ensures that a gene's expression levels can be predicted recursively from its own cis-eQTLs and the cis-eQTLs and/or predicted expression levels of its parent genes in the network. We tested our models on simulated data from the DREAM5 Systems Genetics Challenge \cite{pinna_simulating_2011} and real data from eQTl mapping studies in yeast \cite{albert_genetics_2018} and  human (the Geuvadis study) \cite{lappalainen_transcriptome_2013}. 

\section{Data and Methods}

\begin{figure*}
\centering
\includegraphics[width=0.95\linewidth]{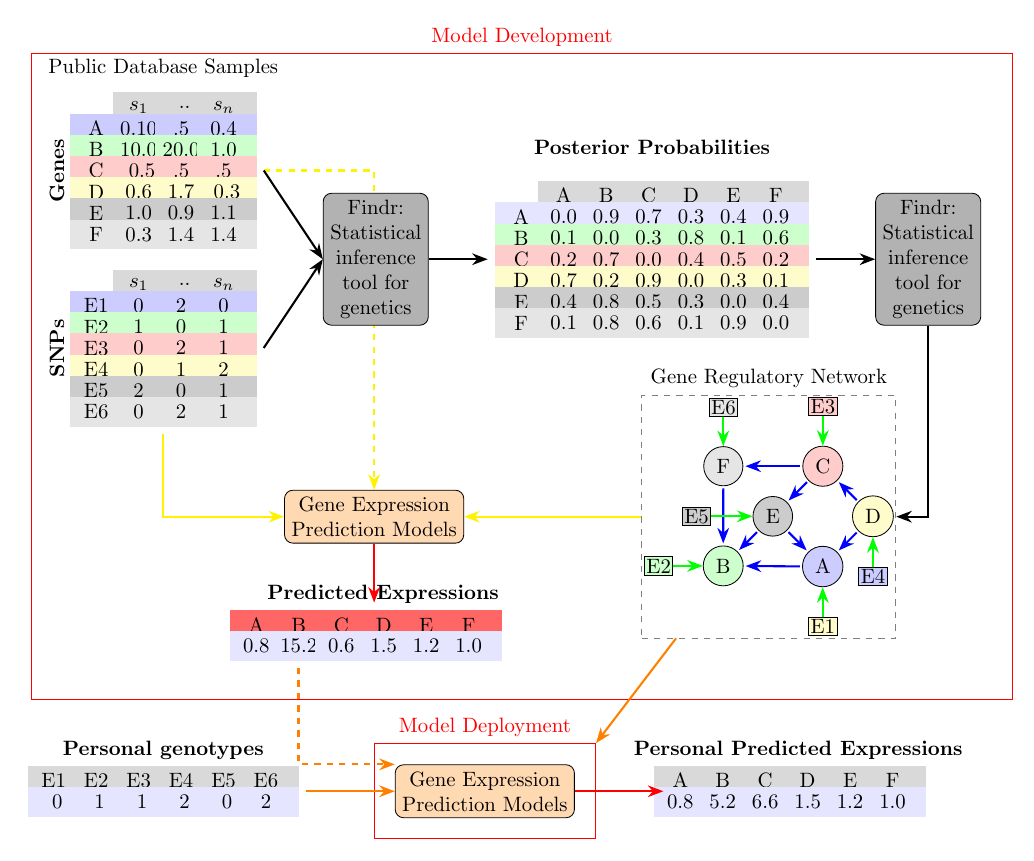}
    \caption{Our pipeline for developing and deploying a prediction model for the genetic component of gene expression levels using a Gene Regulatory Network (GRN) consists of several stages. Initially, we start with the raw gene expression data for a specific set of genes, denoted by letters A to F. Subsequently, we identify the top cis-eQTLs (E1 to E6) that correspond to these genes. The matching colors between the eQTLs and genes indicate their correspondence. Next, we employ the Findr-tool to obtain Bayesian posterior probabilities, which provide an estimate of the likelihood that genes interact with each other. Findr generates posterior probability matrices for five different causal interactions and one correlation interaction. These matrices serve as priors for the network reconstruction method. The stages from the raw data to the GRN reconstruction are represented by black arrow color. On the other hand, the GRN itself is illustrated using blue arrows, and the associated eQTL for each gene is indicated by green arrows. In practice, each gene can be associated with multiple eQTLs. Once the GRN is reconstructed, we can utilize the directed acyclic graph (DAG) structure of the GRN to train a prediction model for the gene expression levels of each gene. This model incorporates the gene's own cis-eQTLs as well as the cis-eQTLs and/or predicted expression levels of its parent genes in the network. During the model development stage, the yellow arrow color is used to indicate the inputs to the model. On the other hand, during model deployment, the orange arrow color is used to represent the inputs. The gene expression and predicted gene expression inputs are represented with dashed lines to indicate that this is an iterative process. }
    \label{fig:gene_snp_matrices}
\end{figure*}

\subsection{Gene expression prediction in TWAS}

Existing methods for transcriptome imputation (TI) in TWAS can be represented as follows \cite{leeuw_interpretation_2023}. The gene expression level $X_g$ of a specific gene $g$ is modelled by the equation: 
\begin{equation}
X_g =  E_g \alpha_g + \xi_g \label{eq:traditional_ti}
\end{equation} 
Here, $E_g$ represents the genotype matrix of cis-eQTLs of gene $g$ (SNPs local to and strongly associated with the gene), and $\xi_g$ is the residual component independent of $E$. In equation (\ref{eq:traditional_ti}), any trans-effect of SNPs elsewhere in the genome is captured by $\xi_g$. This model is trained on a sample dataset with genotype and expression data to estimate the genetic effect vector $\hat{\alpha}_g$. Based on this, the estimated genetic component of gene expression is calculated as \begin{align}
\hat{X}_g^{genetic}  &=  E_g \hat{\alpha}_g \label{eq:trad_ti_ge_comp}
\end{align}

Our TI approach can be described as follows. Consider \( G = (V, \mathcal{E}) \) as a directed acyclic graph (DAG) representing a gene regulatory network (GRN) of \( n \) genes. In this context, \( V = \{g_1, g_2, \ldots, g_n\} \) denotes the set of nodes where each node \( g_i \) corresponds to a specific gene. The set \( \mathcal{E} \) consists of directed edges \( (g_i, g_j) \), where a directed edge from \( g_i \) to \( g_j \) signifies that gene \( g_i \) regulates gene \( g_j \). Let \( X_i \) represent the expression level of gene \( g_i \) within this network. The genotype matrix of the cis-eQTLs for gene \( g_i \) is denoted by \( E_i \), while \( E_i^p \) represents the genotype matrix of the cis-eQTLs for all immediate predecessors (parents) of \( g_i \), identified as the set of nodes \( \{g_k \mid (g_k, g_i) \in \mathcal{E}\} \). Furthermore, \( E_i^{gp} \) denotes the genotype matrix of the cis-eQTLs for ancestors up to grandparents, comprising nodes \( \{g_l \mid \exists (g_l, g_k) \in \mathcal{E} \text{ and } (g_k, g_i) \in \mathcal{E}\} \). Finally, \( X_i^p \) refers to the matrix of expression levels for all parents of \( g_i \).

In our methodology, we distinguish between the cis-genetic component and the trans-genetic component of gene expression. Initially, we model the expression of each gene $g_i$ in the graph $G$ using the traditional method as shown in equation (\ref{eq:traditional_ti}), and then estimate the cis-genetic component: 
\begin{align} \hat{X}_i^{cis} &= E_i \hat{\alpha}_i \label{eq:cis_genetic}\end{align}

If $g_i$ is a root node (a gene with no predecessors or incoming edges), then its genetic component of gene expression is equivalent to the cis genetic component (\ref{eq:cis_genetic}).

For all other genes, we model their trans-genetic component by first calculating the residual (variance of gene $g_i$ not explained by $E_i$) as: 
\begin{equation} 
\xi_i = X_i - \hat{X}_i^{cis} \label{eq:res} 
\end{equation} 

Each residual component $\xi_i$ is then modelled by the equation:
\begin{equation}
    \xi_i = P_i \beta_i + \sigma_i \label{eq:trans_model}
\end{equation}
In equation (\ref{eq:trans_model}), $P_i$ represents the input matrix derived from the parents of $g_i$, $\hat{\beta}_i$ denotes the trans effect vector of $P_i$ on $g_i$, and $\sigma_i$ signifies the noise component.

To obtain the trans-genetic  effect vector  $\hat{\beta}_i$  for all genes, we train equation (\ref{eq:trans_model}) using different parent inputs $P_i$ on sample data (see below):

\begin{itemize}
    \item using cis-eQTL genotypes of parent genes ($E^p$) :
    \begin{align} 
    \hat{X}_i^{trans} &=  \hat{\xi}_i = E_i^{p} \hat{\beta}_i  \label{eq:trans_ep} 
    \end{align}

    \item using cis-eQTL genotypes of parent and grandparent genes ($E^{gp}$) :
    \begin{align} 
    \hat{X}_i^{trans} &=  \hat{\xi}_i = E_i^{gp} \hat{\beta}_i  \label{eq:trans_egp} 
    \end{align}
    \item using recursively predicted expression levels of predecessor genes ($X^{pp}$): 
    \begin{align} 
    \hat{X}_i^{trans} &=   \hat{\xi}_i =  X_i^{pp} \hat{\beta}_i
    \label{eq:trans_xpp} 
    \end{align}
    \item using actual expression levels of parent genes ($X^p$): 
    \begin{align} 
    \hat{X}_i^{trans} &= \hat{\xi}_i =   \hat{X}_i^{p}  \hat{\beta}_i
    \label{eq:trans_xp} 
    \end{align}
\end{itemize} 

We then estimate the expression levels of gene $g_i$ by adding the cis component prediction from equation (\ref{eq:cis_genetic}) and one of the trans component predictions from equations \eqref{eq:trans_ep}--\eqref{eq:trans_xp}:

\begin{align}
    \hat{X}_i^{genetic} &= \hat{X}_i^{cis} + \hat{X}_i^{trans} \label{eq:cis_trans_genetic}
\end{align}

The $X^p$ (\ref{eq:trans_xp}) method  represents the optimal prediction model for trans component based on the trained models, but it is not practical for TI because it assumes  parent expression data. It is included solely for benchmarking purposes. 

The $X^{pp}$ (\ref{eq:trans_xpp}) method is the theoretical approach to predicting the expected expression levels of all genes in the joint SNP and gene Bayesian network model, given observed genotype values. Starting from root nodes where the genetic component equals to cis component we can recursivly predict parent expression levels $X^{pp}$  using equation (\ref{eq:trans_xp})  and (\ref{eq:cis_trans_genetic}). If the GRN $G$ were the true GRN, this method would yield the true genetic component of gene expression. However, with an imperfect GRN, the recursive estimator might propagate errors.

The $E^{p}$ (\ref{eq:trans_ep}) and $E^{gp}$ (\ref{eq:trans_egp}) methods represent a compromise strategy where recursive gene expression prediction is halted after one and two steps, respectively. In other words, we substitute the parent expression with its estimator based on the traditional TI model \eqref{eq:trad_ti_ge_comp} and disregard the indirect effects of SNPs that are cis-eQTLs to ancestors beyond the grand parent genes.

A graphical summary and detailed implementation algorithm of our approach are given in Figure \ref{fig:gene_snp_matrices} and  SI Section \ref{main_algorithm}, respectively.

\subsection{Regression models} 

Similar to existing transcriptome imputation methods, we trained ridge regression (L2 regularization), lasso regression (L1 regularization), elastic net regression (L1 and L2 regularization) \cite{hastie2009elements}, and Bayesian ridge regression models to predict gene expression levels based on genetic information. All methods were implemented using the Python scikit-learn library \cite{pedregosa_scikit-learn_2011}. Both the lasso and ridge models were optimized using cross-validation estimators. The alpha values for both models ranged from 0.001 to 10, with a total of 50 values. Similarly, the elastic net model was also optimized using cross-validation estimators, using the same range of alpha values. In addition, the elastic net model used l1 ratio values of 0.01, 0.5, and 1.0. On the other hand, the Bayesian ridge model was trained with default parameters, except for the max iteration set to 10,000. All models were evaluated using 5-fold cross-validation.

\subsection{Data}
To evaluate the effectiveness of gene expression prediction techniques, we utilized data from a DREAM challenge, a yeast eQTL mappnig study and the Geuvadis study. 

The DREAM5 Systems Genetics Challenge A, also referred to as DREAM, utilized the SysGenSIM software \cite{pinna_simulating_2011} to generate synthetic genotype and transcriptome data for artificial gene regulatory networks. For our analysis, we utilized the 1000 gene network and 1000 sample dataset. We restricted our analysis to the 250 genes in the network with cis-eQTLs. The dataset is available at \href{https://www.synapse.org/}{https://www.synapse.org/}. Further details about the data preprocessing are in the Supplementary Information.

We used genotype and transcriptome sequencing data from a study where two yeast strains, a laboratory strain (BY) and a wine strain (RM), were crossed to generate 1,012 segregants (samples)\cite{albert_genetics_2018}.  The study assessed mRNA expression levels for 5720 genes and genotypes for 11,530 variant sites in each segregant, and identified 2969 local eQTLs that affected 2884 genes. We restricted our analysis to these 2884 genes with cis-eQTLs. Similar to the original study, we corrected gene expression levels for batch effect and optical density during culture growth. The dataset can be accessed at \href{https://figshare.com/s/83bddc1ddf3f97108ad4}{https://figshare.com/s/83bddc1ddf3f97108ad4}

We used genotype and RNA-sequencing data from lymphoblastoid cell lines (LCLs) from 362 individuals of European descent in the Genetic European Variation in Disease (Geuvadis) study \cite{lappalainen_transcriptome_2013}. The eQTL discovery analysis carried out by the Geuvadis project identified local cis-eQTLs for 3,285 genes at a false discovery rate (FDR) of 5\%. We restricted our analysis to these 3285 genes with cis-eQTLs. The dataset can be accessed at \href{https://ftp.ebi.ac.uk/pub/databases/microarray/data/experiment/GEUV/E-GEUV-1/}{https://ftp.ebi.ac.uk/}. Further details about the data preprocessing are in the Supplementary Information.

For all datasets, we randomly divided the samples into training (80\%) and test sets (20\%), ensuring that gene expression and SNP genotype samples remain aligned in the random sampling. The gene regulatory network and gene expression prediction models are then reconstructed solely from the training data, while the test data is used solely for evaluating and validating the TI model.

The DREAM and yeast  data had approximately 800 training samples and 200 test samples, while Geuvadis had 270 and 70  training and test samples, respectively. For DREAM and yeast, each gene had only one eQTL associated with it, but for Geuvadis, each gene had hundreds of eQTL associations. We present only the results of using one eQTL per gene here; the results of using all eQTL can be found in Supplementary Information \ref {snp_analysis}

 In the case of models that used both genotype and expression features, we standardized the expression features by subtracting the mean and dividing by the standard deviation before concatenating them. The scaling method was fitted on the training data and applied to the testing data. During testing, the predicted expression features were scaled  using the training scalers.

 \subsection{GRN reconstruction}

We utilized the software package Findr \cite{wang_efficient_2017} to reconstruct GRNs from systems genetics data. More specifically, we employed the ``pij\_gassist'' function, which takes gene expression levels (quantified as RNA transcript read count) and the genotypes of the strongest cis-eQTL of all or a subset of genes as input. For each pair of genes, the function conducts six likelihood ratio tests. It returns for each test the Bayesian posterior probability of the appropriate (alternative or null) hypothesis being true. These posterior probabilities are equivalent to the complement of the local false discovery rate (lfdr) ($P=1 - lfdr$) \cite{hastie2009elements}. For a gene $A$ with strongest cis-eQTL $E_A$ and putative target gene $B$, the six likelihood ratio tests or posterior probabilities are: $P_0$:  $B$ is correlated with $A$ (``correlation''); $P_1$: $E_A$ is associated with $A$ (``primary linkage''); $P_2$: $E_A$ is associated with $B$ (``secondary linkage''), $P_3$: $B$ is independent of $E_A$ given $A$ (``conditional independence''), $P_4$: $B$ is not independent of $E$ and $A$ given the association $E_A\to A$ (``relevance''), and $P_5$: $B$ is not independent of $A$ given $E_A$ (``controlled''). 

The six probabilities are then combined in different ways to estimate the probability $P(A\to B)$ of a direct edge from $A$ to $B$, leading to multiple GRNs: the  correlation GRN ($P_0$), the mediation GRN ($P_2P_3$), the instrumental variable GRN ($P_2P_5$), and a combination that best accounted for hidden confounders ($P=0.5(P_2P_5 + P_4)$). Note that due to considering only genes with a significant eQTL, we always have $P_1=1$ and this factor is therefore not written explicitly in these formulae. From the final probability matrices, we retained only the square matrices of pairs where both genes have cis-eQTLs (``A''-genes in Findr terminology).

To construct a DAG from a given square probability matrix, we used the "netr\_one\_greedy" function in Findr, which adds edges one by one by decreasing order of their probability, skipping edges that would introduce a cycle among the already added nodes \cite{wang_high-dimensional_2019}. The function generates a boolean value for each edge indicating the presence or absence of a directed edge between each pair of genes in the DAG.

We only kept the edges in the final GRNs that had a probability exceeding a certain threshold. The thresholds were set to ensure that different networks had similar characteristics, such as total nodes, edges, root, intermediate, and leaf nodes. We calculated the global false discovery rate (FDR) as one minus the average posterior probability of the interactions we kept (\cite{chen2007harnessing}). The full overview of all networks and their characteristics can be found in table \ref{tab:network_statistics_comparison} in the Supplementary Information.

We also generated randomized GRNs by performing a topological sort on the original DAG of the GRN, followed by shuffling this order to create a new node sequence. Nodes are then relabeled according to this new sequence. This approach preserves the original network's acyclicity and connectivity patterns, maintaining the hierarchical relationships crucial for our goal of predicting each gene's expression level based on its parent and grandparent information. While we considered alternative randomization methods, such as degree-preserving randomization, these approaches often introduce computational complexities and can compromise the acyclic nature of the network. The acyclic property is essential for our hierarchical prediction framework, as it allows us to accurately model the dependencies and causal relationships between genes.

\section{Results}\label{sec4}

We aim to develop a prediction model for each gene within a GRN to predict the genetic component of gene expression level variation, which can be used for transcriptome imputation (TI) in transcriptome-wide association studies. Each predicted genetic component of a gene is the sum of two model predictions: the cis-component prediction model and the trans-component prediction model. The cis-component model, which we refer to as ``\textbf{$E$}", is a traditional TI model that uses only the genotypes of a gene's cis-eQTLs as predictors. For the trans-component model, we experimented with four different inputs. ``$E^{p}$", which considers the cis-eQTLs of a gene's parent genes in the GRN as predictors. ``$E^{gp}$", which considers the cis-eQTLs of both a gene's parent and grandparent genes in the GRN as predictors. ``$X^{p}$", which uses the actual expression of its parent genes in the GRN as predictors; and ``$X^{pp}$", which uses the recursively predicted expression of its parent genes in the GRN as predictors. Note that ``$X^{pp}$" is not an independent model; instead, it is the same model as $X^{p}$, meaning the model is fitted on actual expression data of parent genes, but the actual data of parent genes is replaced by a recursively predicted expression of parent genes during inference. In the recursive approach, root nodes (genes without parents in the GRN) genetic component are predicted only from their own cis-eQTLs. These predicted expression levels are then used to predict the trans component of expression levels of child nodes of the root nodes, which is combined with its cis component to give a full genetic component prediction. This process is iterated until all leaf nodes (genes without descendants in the GRN) are reached (see Methods for details).

For our results, we only considered models of intermediate and leaf nodes in the GRN. We also introduced a predictability criterion where a gene is considered predictable if it achieves an $R^2$ value of at least $0.05$ using the combined input of cis-eQTLs and parent expression $(E + X^p)$. This threshold aligns with definitions used in previous studies, such as those by \cite{zeng_prediction_2017}, who defined genes as predictable with an $R^2$ value of at least $0.05$, and \cite{basu_predicting_2021}, who used similar criteria for defining predictability and high predictability in gene expression models.

This ensures that only genes with predictable genetic components are included in our analysis. At the same time it also includes genes that are not well predictable from the baseline model but predictable when adding parental information.

Out of the total genes analyzed, a significant proportion met the predictability criterion, achieving an $R^2$ value $\geq 0.05$ across all the  models (Lasso, Ridge, Bridge, and Elastic Net) and data (DREAM, yeast and Geuvadis). We also observed that while the mean $R^2$ and the number of selected nodes varied with different $R^2$ thresholds, the overall performance pattern remained consistent.

\subsection{Comparison of prediction methods}

We used four regularized regression models: Lasso, Ridge, Bayesian Ridge, and Elastic Net, to predict gene expression levels. These models were evaluated using five different inputs: traditional model using only cis-eQTLs as predictors ($E$), model using own cis-eQTLs and parent eQTLs ($E + E^p$), model using own cis-eQTLs , parent cis-eQTLs, and grandparent cis-eQTLs ($E + E^{gp}$), model using own cis-eQTLs and actual parent gene expression levels ( $E + X^p$), model using own cis-eQTLs and recursively predicted parent expression levels ($E + X^{pp}$).

The performance of these models was evaluated using the $R^2$ score, which measures how well the independent variables explain the variance in the target variable (gene expression level). The results are summarized in  Table \ref{table:model-comp-results} and Figure S1.

\begin{table}
\centering
\caption{Mean $R^2$ values for each combination of dataset, input type, and regularized regression model. Bold values indicate the best performance for a practically relevant model, while italic values represent the theoretical optimum but are considered irrelevant for practical TI.}
\label{table:model-comp-results}
\begin{tabular}{llcccc}
\toprule
\textbf{Dataset} & \textbf{Input} & \textbf{Lasso} & \textbf{Ridge} & \textbf{Bridge} & \textbf{ElNet} \\
\midrule
\multirow{5}{*}{DREAM}    & $E$            & 0.109 & 0.108 & 0.109 & 0.109      \\
                          & $E + E^p$       & 0.179 & 0.178 & 0.179 & 0.178       \\
                          & $E + E^{gp}$      & \textbf{0.191} & \textbf{0.185} & \textbf{0.188} & \textbf{0.190}       \\
                          & $E + X^{pp}$     & 0.121 & 0.120 & 0.121 & 0.121       \\
                          & $E + X^p$       & \emph{0.223} & \emph{0.223} & \emph{0.223} & \emph{0.223}       \\
\midrule
\multirow{5}{*}{Yeast}    & $E$            & 0.126 & 0.126 &  0.126 &  0.126       \\
                          & $E + E^p $      & 0.214 & 0.212 & 0.213 & 0.214       \\
                          & $E + E^{gp}$      & \textbf{0.233} & \textbf{0.227} & \textbf{0.232} & \textbf{0.233}       \\
                          & $E + X^{pp}$     & 0.176 & 0.176 & 0.176 & 0.176       \\
                          & $E + X^p$       & \emph{0.487} & \emph{0.488} & \emph{0.488} & \emph{0.487}       \\
\midrule
\multirow{5}{*}{Geuvadis} & $E $           & 0.094 & 0.095 & 0.095 & 0.095       \\
                          & $E + E^p$       & \textbf{0.119} & \textbf{0.119} & \textbf{0.120} & \textbf{0.119}       \\
                          & $E + E^{gp}$      & 0.100 & 0.090 & 0.099 & 0.096       \\
                          & $E + X^{pp}$      & 0.108 & 0.108 & 0.108 & 0.108       \\
                          & $E + X^p $      & \emph{0.232} & \emph{0.232} &\emph{0.232} & \emph{0.232}       \\
                          
\bottomrule
\end{tabular}

\end{table} 

In all datasets, the incorporation of parent cis-eQTLs ($E + E^p$) consistently enhances the mean $R^2$ values compared to the traditional cis-eQTLs model (E). The inclusion of grandparent eQTLs ($E + E^{gp}$) generally leads to further improvements, although the degree of enhancement varies across datasets. For example, the mean $R^2$ significantly increases in the DREAM dataset by including grandparent eQTLs. However, in the Geuvadis dataset, this approach has a negative effect, with all models showing a decrease in performance when adding grandparent eQTLs.

Utilizing the Actual Parent Gene Expression $(E + X^p)$  model unsurprisingly produces the highest average $ R^2$ values across all datasets and regression techniques. These results support our hypothesis that most expression variance is attributed to trans-effects propagated through the networks, and having true parental expression will contain all information to be propagated.  The performance of the Recursively Predicted Parent Expression ($E + X^{pp}$) model is surprisingly low in all datasets datasets.  All models utilizing parent eQTLs, $E + E^{p}$ has better performance than $E + X^{pp}$.

\subsection{Bayesian Ridge Regression on different inputs}

\begin{figure}
    \centering
    \textbf{A. DREAM}
    \vspace*{-2mm}
    \begin{center}
        \includegraphics[width=.7\linewidth]{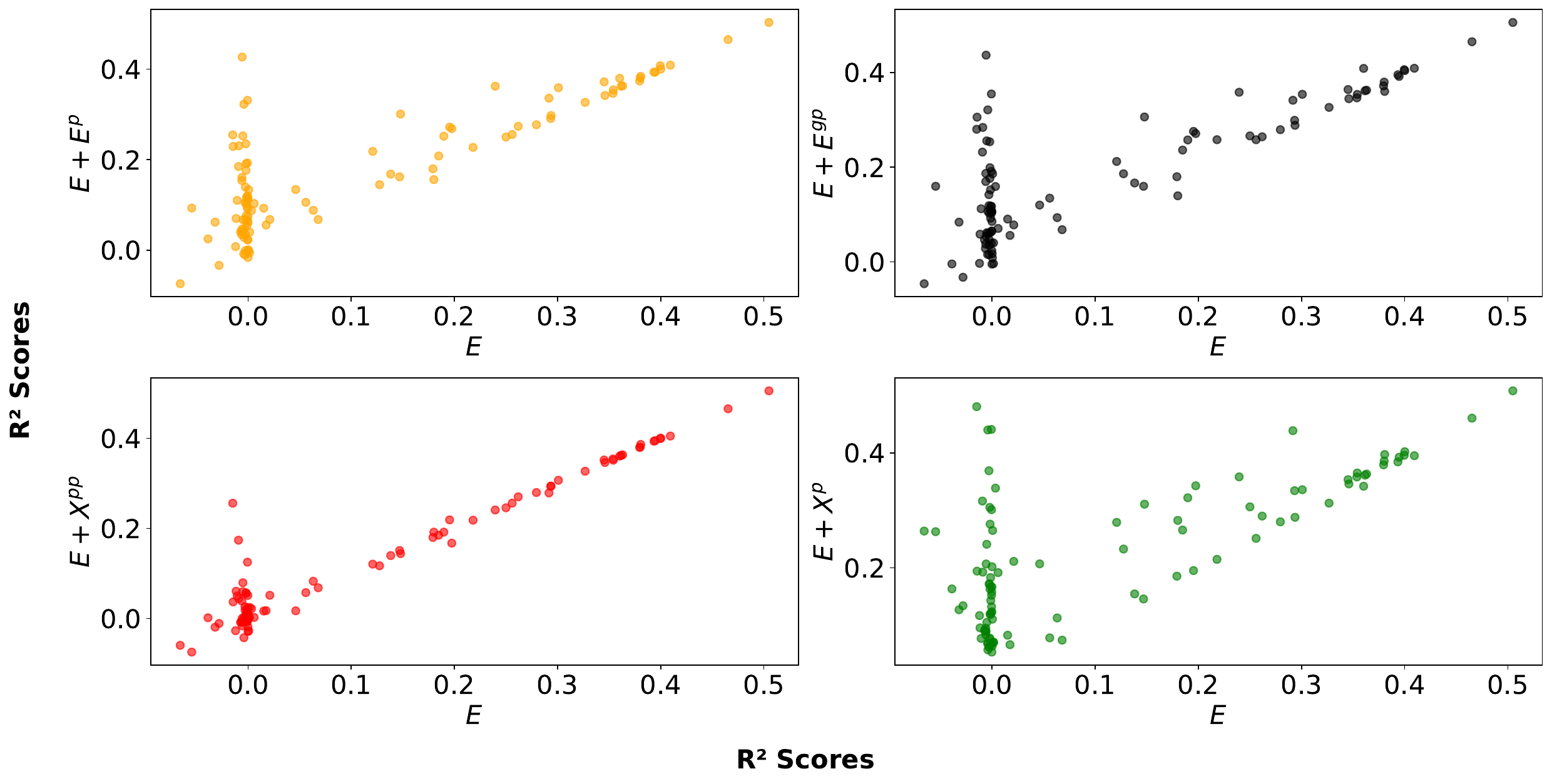}
    \end{center}
    
    \textbf{B. Yeast}
    \vspace*{-2mm}
    \begin{center}
        \includegraphics[width=.7\linewidth]{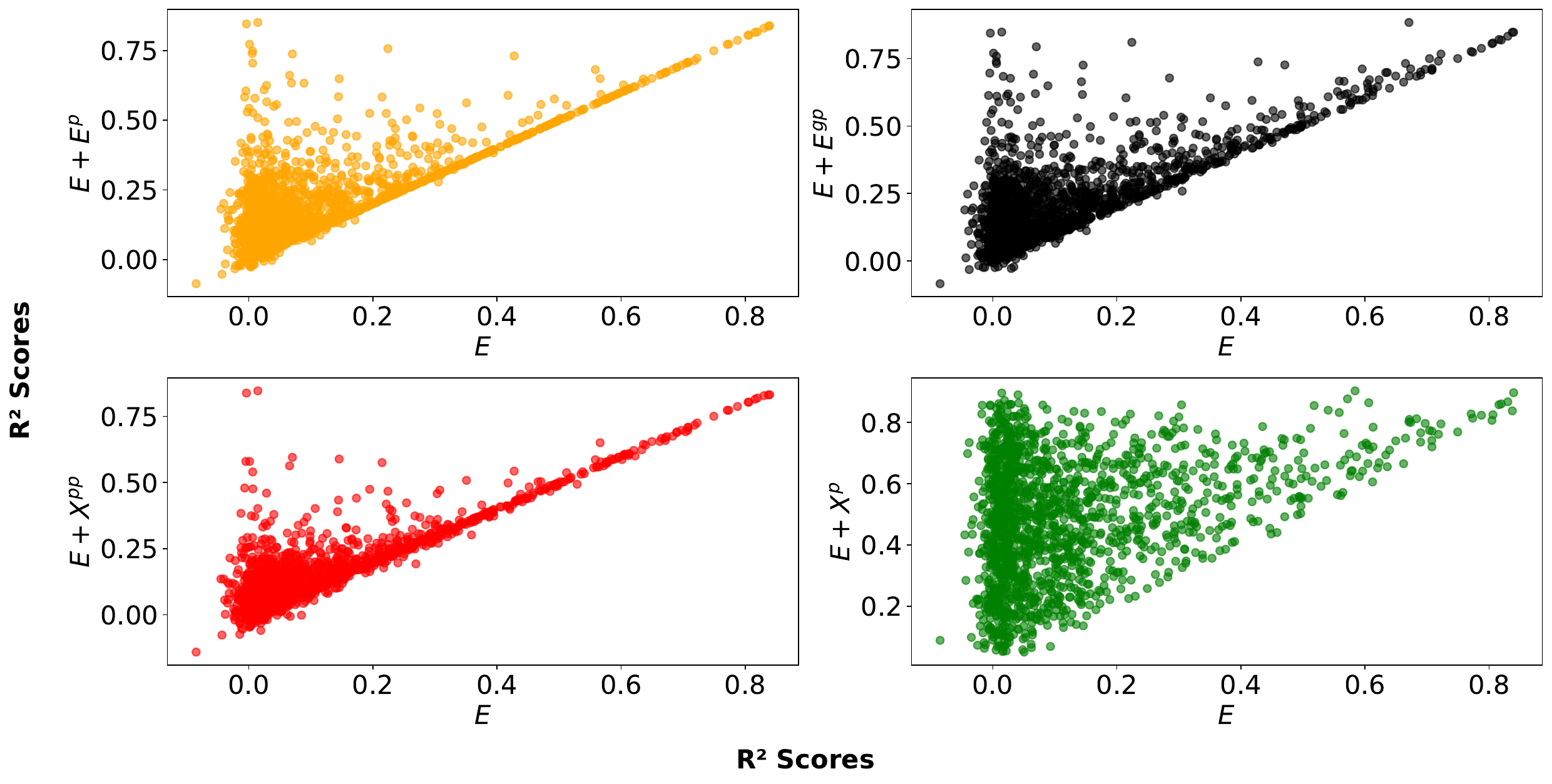}
    \end{center} 
    \textbf{C. Geuvadis}
    \vspace*{-2mm} 

      \begin{center}
        \includegraphics[width=.7\linewidth]{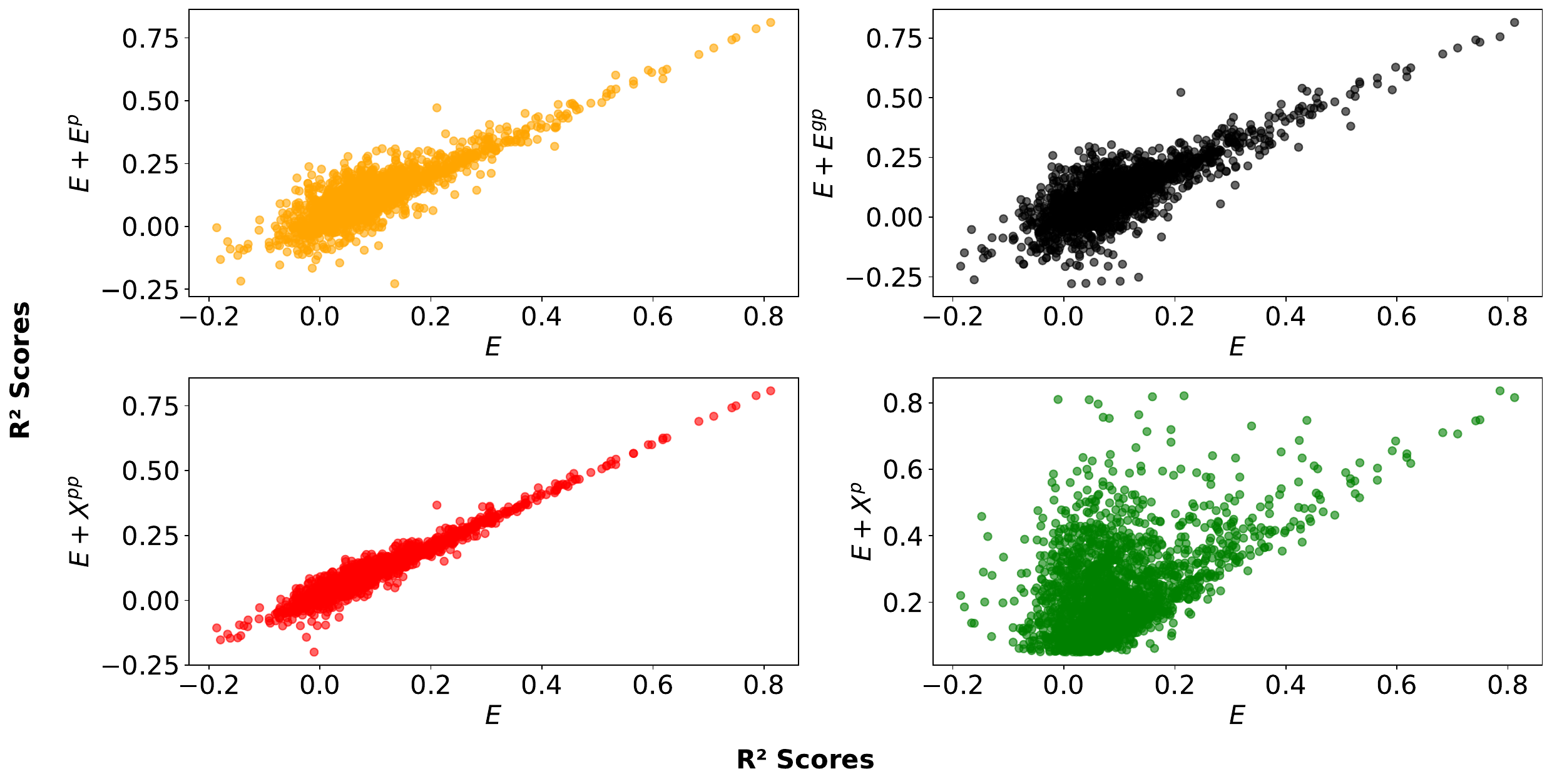}
    \end{center} 
  \caption{Scatter plots comparing $R^2$ values for each gene predicted by the traditional cis-eQTL model (E) on the x-axis with models incorporating parent information on the y-axis across three datasets: A. DREAM, B. Yeast, and C. Geuvadis.}

    \label{fig:compare_inputs_pairgrid}
\end{figure}

To compare different input methods further, we used the Bayesian ridge regression models trained using the Findr-$P$ causal GRN reconstructed from the DREAM, yeast, and Geuvadis data.

Figure \ref{fig:compare_inputs_pairgrid} displays the $R^2$ value for each gene in the network. It compares model performance across the dataset, emphasizing the differences in individual gene prediction when using parent information (y-axis) compared to the traditional method (x-axis) that only utilizes cis-eQTLs. 

The overall trend in Figure \ref{fig:compare_inputs_pairgrid} is encouraging. It suggests that most genes, which initially had low predictability ($R^2 < 0.30$) with the baseline method, show a significant improvement in prediction accuracy when we incorporate parental information. This enhancement is consistent across all datasets. We also observe this trend with recursively predicted parental expression inputs in large sample datasets (DREAM and yeast). However, in the Geuvadis dataset, the patterns are not as clear, except for the actual parent expression.

\subsection{Relevance of network information}

To analyze the impact of various network types, we utilized the Findr method to reconstruct two causal networks ($P$ and $P_2P_5$) and one correlation network ($P_0$). In addition, we generated five random networks by shuffling the nodes in $P$. This approach enabled us to evaluate the performance differences arising from randomness or the addition of informative networks (See Method section for details).

Given that each true network is naturally distinct, direct comparison is not possible without permitting some networks to have more false connections (higher global FDR, see Methods). In this study, we adjusted the edge probability thresholds to ensure that the networks exhibit comparable characteristics such as the number of nodes and edges (Table \ref{tab:network_statistics_comparison}). Keeping these characteristics constant ensured that prediction performance is compared across a comparable number of genes and that on average, the number of parent nodes per gene is comparable.

With the linear regression methods showing a comparable average performance on the Findr-$P$ network (Figure \ref{fig:compare_model}), we turned our attention to the application of Bayesian ridge regression to the three real networks and five random networks. The following analysis of the average $R^2$ performance for our five inputs across each network, utilizing simulated DREAM data, yeast, and Geuvadis data (Fig. \ref{fig:compare_network_mean_bar}A, B, and C, respectively), uncovered several observations. Note that the result for the random network represents the average of five distinct random networks.

Across all datasets (Table \ref{table:network-comp-results}, and Fig. \ref{fig:compare_network_mean_bar}), causal networks ($P$ and $P_2P_5$) generally outperforms the correlation network ($P_0$) and random networks when predicting gene expression from eQTL data. The best performance is typically observed with the $P$ network indicating the superior ability to capture underlying causal relationship in gene expression regulation. The worse performance of the correlation network $P_0$ is expected given correlation does not imply causation and may include spurious relationships. Random networks, show the lowest performance, underscoring the credibility of the real networks and indicating the structured, biologically-informed connections in the real networks ($P,P_2P_5$ and $P_0$) captures meaningfull regulatory relationship that are crucial for accurate gene expression prediction. 

\begin{table}
\centering
\caption{Mean $R^2$ values for each combination of dataset, input type, and networks. Bold values indicate the best performance for a practically relevant model, while italic values represent the theoretical optimum but are considered irrelevant for practical TI.}
\label{table:network-comp-results}
\begin{tabular}{llcccc}
\toprule
\textbf{Dataset} & \textbf{Input} & \textbf{$P$} & \textbf{$P_2P_5$} & \textbf{$P_0$} & \textbf{Random} \\
\midrule
\multirow{5}{*}{DREAM}    & $E$            & 0.13 & 0.13 & 0.13 & 0.13      \\
                          & $E + E^p$       & 0.20 & 0.19 & 0.15 & 0.13       \\
                          & $E + E^{gp}$      & \textbf{0.21} & \textbf{0.20} & \textbf{0.16} & \textbf{0.14}       \\
                          & $E + X^{pp}$     & 0.14 & 0.14 & 0.13 & 0.13       \\
                          & $E + X^p$       & \emph{0.24} & \emph{0.23} & \emph{0.26} & \emph{0.14}       \\
\midrule
\multirow{5}{*}{Yeast}    & $E$            & 0.11 & 0.11 &  0.11 &  0.11       \\
                          & $E + E^p $      & 0.21 & 0.21 & 0.20 & 0.13       \\
                          & $E + E^{gp}$      & \textbf{0.23} & \textbf{0.23} & \textbf{0.24} & \textbf{0.16}       \\
                          & $E + X^{pp}$     & 0.17 & 0.17 & 0.12 & 0.11       \\
                          & $E + X^p$       & \emph{0.50} & \emph{0.50} & \emph{0.65} & \emph{0.35}       \\
\midrule
\multirow{5}{*}{Geuvadis} & $E $           & 0.13 & 0.13 & 0.13 & 0.13       \\
                          & $E + E^p$       & \textbf{0.16} & \textbf{0.15} & \textbf{0.13} & \textbf{0.12}       \\
                          & $E + E^{gp}$      & 0.15 & \textbf{0.15} & 0.11 & 0.09      \\
                          & $E + X^{pp}$      & 0.14 & 0.14 & 0.13 & 0.12       \\
                          & $E + X^p $      & \emph{0.26} & \emph{0.23} &\emph{0.32} & \emph{0.25}       \\
                          
\bottomrule
\end{tabular}

\end{table}

In all datasets, the performance of the correlation network $P_0$ is better than all other networks when true parent expression levels are used.

We believe this is because true parent expression levels provide more precise information about gene expression, thereby reducing reliance on the quality of the network structure and allowing even correlation networks to perform well. Since correlation networks capture both true and spurious relationships regarding parental eQTLs, their predictive power diminishes in comparison to causal networks. Similarly, the recursive prediction method propagates compound errors from previous predictions in correlation networks, These compound errors are exacerbated due to the inclusion of non-causal connections. Furthermore, when using small sample sizes, which already suffer from overfitting and variability issues, the spurious connections in correlation networks can significantly impact model performance. This is evident in the performance of the correlation network when using Geuvadis data (Figure \ref{fig:compare_network_mean_bar} C).

One particularly noteworthy finding is the performance of the recursively predicted inputs $E + X^{pp}$  which shows a slight improvement over the baseline method $E$ in causal networks, and performs less than baseline in correlation and random networks across all datasets. This underscores the crucial role of accurate network structure. In the context of causal networks, the recursive prediction leverages more meaningful biological relationships, thereby enabling the model to capture additional regulatory information and enhance the prediction accuracy. Conversely, the absence of a genuine regulatory structure in correlation and random networks results in compounded error and inferior performance.

We  observed that the performance of random networks on the yeast data is unexpectedly high for the $E + X^p$ model (Fig. \ref{fig:compare_network_mean_bar}). Upon investigation, we concluded that this dataset exhibits a very high correlation among all genes (see SI Section \ref{correlation_analysis}). Consequently, even randomly selected parents can show a high correlation with the gene of interest.

\begin{figure}
    \textbf{A. DREAM}
    \vspace*{-2mm}
    \begin{center}
    \includegraphics[width = .45\linewidth]{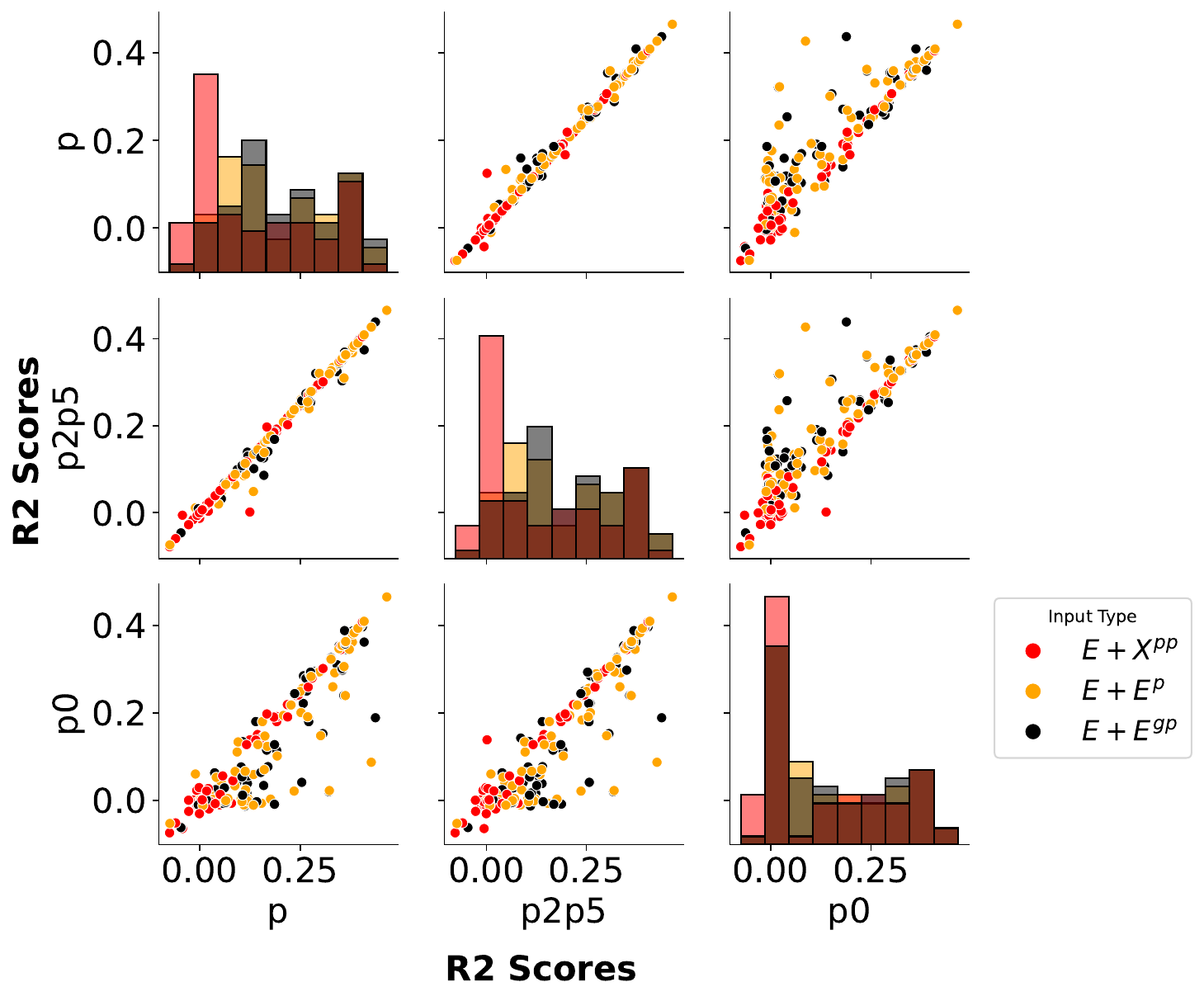}
    \end{center}
    
    \textbf{B. Yeast}
    \vspace*{-2mm}
    \begin{center}
         \includegraphics[width = .45\linewidth]{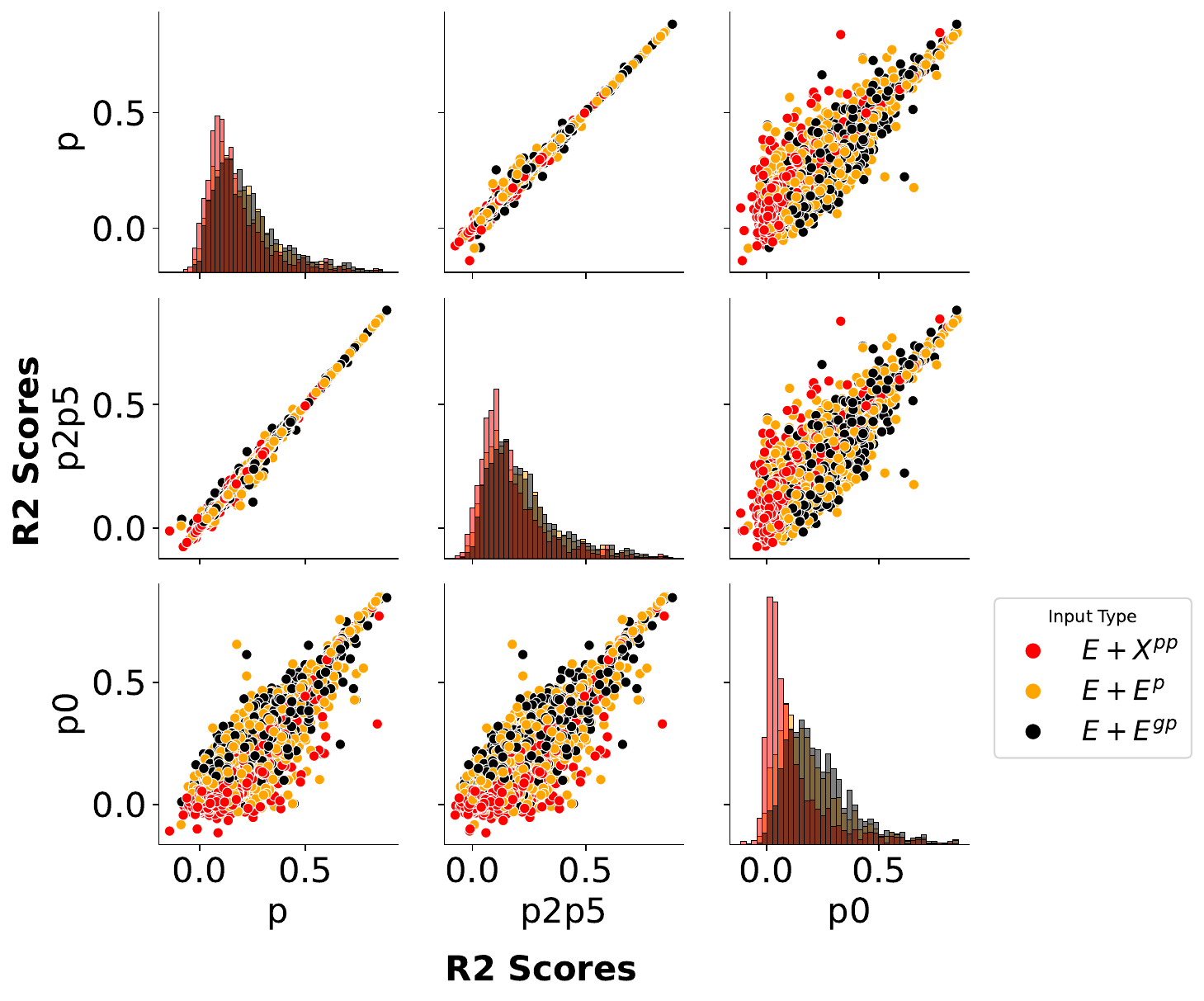}
    \end{center}

 \textbf{C. Geuvadis}
    \vspace*{-2mm}
    \begin{center}
         \includegraphics[width = .45\linewidth]{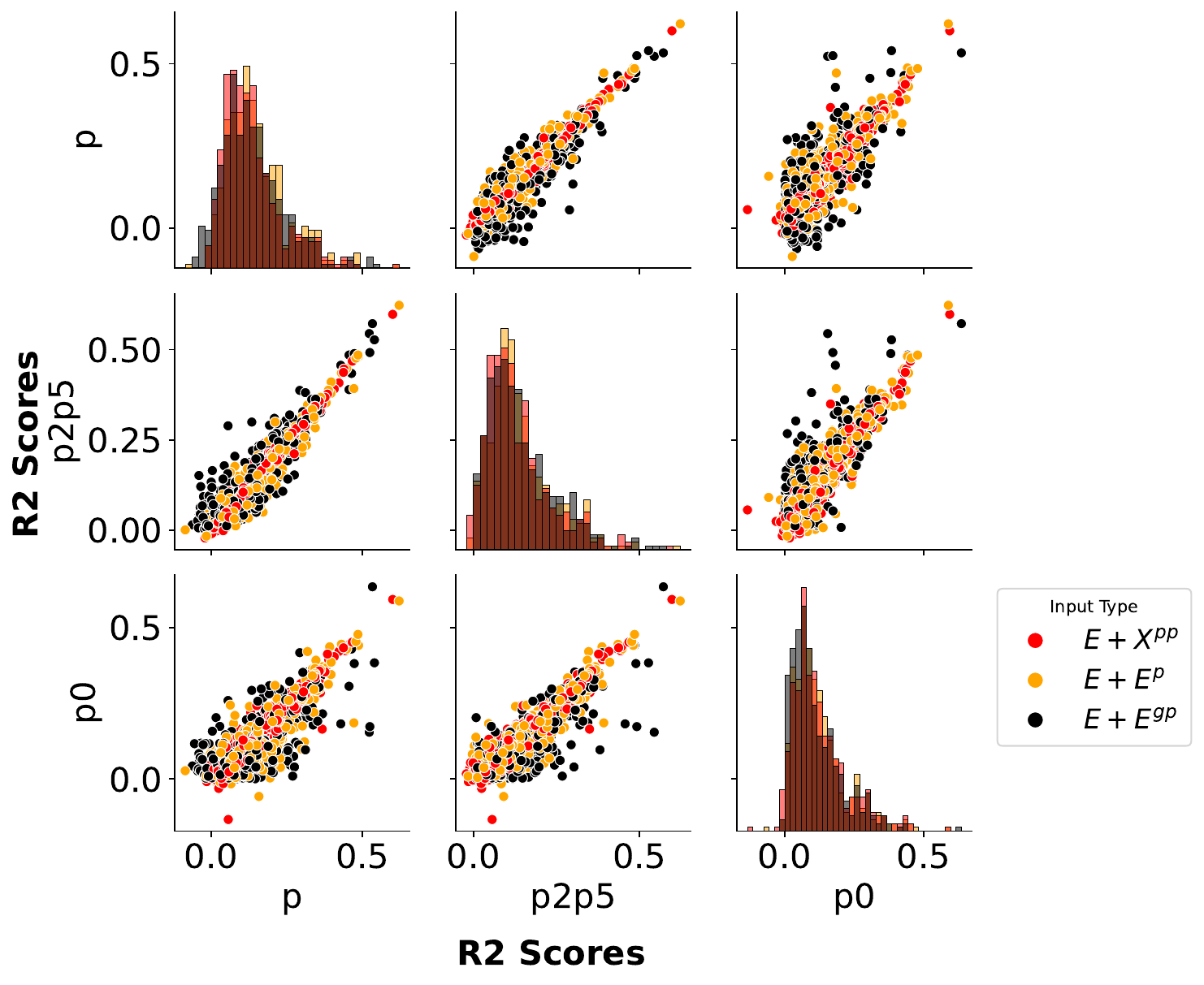}
    \end{center}
    
      \caption{Pair grid results comparing $R^2$ values for each gene predicted by different reconstructed networks across three datasets: A. DREAM, B. Yeast, and C. Geuvadis. Each subplot in the grid compares the performance of pairs of networks ($P$, $P_2P_5$, $P_0$) for three input combinations: $E + E^p$, $E + E^{gp}$, and $E + X^{pp}$.}
    \label{fig:compare_network_pair_grid}
\end{figure}

To obtain a more detailed understanding of how the model behaves when different networks are used, we compared the performance to predict the expression of individual genes across all pairs of real networks and three input combinations focusing on three three input combinations that does not use actual parent expressions $E + E^p$, E$ + E^{gp}$, and $E + X^{pp}$ (Figure \ref{fig:compare_network_pair_grid}).

Our investigation of the  DREAM  and yeast data (Fig \ref {fig:compare_network_pair_grid}A and B) shows that both the causal networks $P$ and $P_2P_5$ demonstrated a linear relationship in their performances on all inputs. This relationship was particularly strong on highly predictable genes ($R^2 > 0.3$), with $P$ showing a slight advantage on genes with less predictability. Importantly, this indicates that both approaches are equally effective in predicting genes when their predictability is high.

Causal networks surpass correlation networks across all datasets and inputs, particularly for $E + E^p$ and $E + E^{gp}$. This is especially evident with the low sample size Geuvadis data (Figure \ref{fig:compare_network_pair_grid} C).

\section{Discussion and Conclusion}\label{sec5}

We presented a new model and method for predicting the genetic component of gene expression in transcriptome-wide association studies (TWAS), based on the model of omnigenic inheritance. The omnigenic model suggests that the genetic component of gene expression is influenced by both cis- and trans-acting variants. In particular, trans-acting variants act through densely connected GRNs. They have small effects that are collectively significant but may be individually insignificant. 

Mathematically, we model the GRN that drives omnigenic inheritance in a certain cell type or tissue as a Bayesian network (BN) jointly over SNPs and genes, where genes are connected by a DAG and SNPs can only directly influence genes as cis-eQTLs and have no incoming edges. We reconstructed GRNs using established coexpression and causal inference methods implemented in the Findr software, where causality is determined from both coexpression and trans-eQTL associations using combinations of likelihood ratio tests.

It is important to note that while a DAG representation offers important computational benefits, real GRNs often include feedback loops and cycles that are not captured by DAGs. However, resolving the cyclic effects of GRNs requires time series data and is not possible using the population-based snapshot data considered here. The DAG reconstruction algorithm used here ensures that if cyclic causal dependencies are predicted from the data, the edges with the strongest genetic support are retained, in line with our aim of predicting the genetic component of gene expression.  

We separate the inference of each gene's expression level into cis and trans components. The cis component is predicted using the standard method, which only uses cis-eQTLs as predictive features (E). The trans component is predicted through the residual of the standard method using parental information as inputs. Our primary residual modeling method is truncating a recursive prediction procedure after one step or two steps. That is, using a gene's parent cis-eQTLs $(E^p)$ as predictive features for the residual model or combining genes' parent and grandparent cis-eQTLs ($E^{gp}$) as a predictive model for the residual. We also added a theoretically optimal artificial method where we assume parent gene expression levels $X^p$ are fully observed to predict the trans component. This results in the best possible prediction score for each gene. We also evaluated a realistic version of the artificial model where $X^p$ of each gene is replaced by recursively predicted $X^{pp}$ values. For each gene, the cis component and trans component prediction are added together to make the predicted genetic component of that gene.

The various prediction methods performed as expected on data simulated from synthetic gene networks. On average, the prediction performance of the GRN-based model was significantly better than the baseline approach and not much below the theoretical optimum with observed parent expression levels. The only surprise was the relatively poor performance of the truncated recursive prediction method, in which the improvement over the baseline method is similar to what one expects from random inputs.

Similar results were observed in data from yeast, where the prediction performance of the GRN-based method that uses parental eQTL information and causal $P$ and $P_2$ methods is almost twice as good as that of the baseline method that uses only its own eQTLs. The performance of the recursive method is also surprisingly higher than the baseline for this dataset compared to the DREAM datasets. 

On real data from lymphoblastoid cell lines from the Geuvadis study similar trends were observed. However, the GRN-based methods showed only a modest improvement over the standard method with greater gap to the theoretical optimum with observed parent expression levels. Adding grandparent eQTLs $E + E^{gp}$ leads to worthier result than the $E + E^p$ model.

The performance of the Recursively Predicted Parent Expression ($E + X^{pp}$) model is surprisingly low in all datasets. The average $R^2$ values for $E + X^{pp}$ are similar to the baseline model. All models utilizing parent eQTLs, $E + E^{p}$, have better performance than $E + X^{pp}$. One potential reason might be that recursively predicting parent expression could introduce additional noise or compounded errors, which could potentially offset the benefits of capturing more complex dependencies, especially if the initial predictions are not highly accurate.

Across all datasets,  causal networks ($P$ and $P_2P_5$) demonstrate superiority over the correlation network ($P_0$) and random networks. The consistent best performance of the $P$ network is a testament to its superior ability to capture underlying causal relationships in gene expression regulation. The underperformance of the correlation network $P_0$, which is expected due to the lack of causation implication and potential inclusion of spurious relationships, further reinforces the confidence in our research methodology. The lowest performance of random networks serves as a contrast, highlighting the structured, biologically informed connections in the real networks ($P,P_2P_5$, and $P_0$) that capture meaningful regulatory relationships crucial for accurate gene expression prediction.  

Going forward, it will be important to repeat our analyses using datasets with larger sample sizes and to compare GRN-based transcriptome imputation to the standard approach in the downstream TWAS steps where gene expression levels are imputed into GWAS and gene-trait associations are analyzed.

While larger eQTL datasets certainly exist, the Geuvadis data has the significant advantage of being one of the very few fully open access datasets. Its genotype data are available with no need to go through a lengthy controlled data access application, thus facilitating rapid and easily reproducible method development. 

For now, causal GRN-based transcriptome imputation requires access to individual-level data. This is not a strict requirement of the Findr network reconstruction method, but rather due to the fact that the necessary summary statistics for causal network reconstruction (trans-eQTL effect sizes) are not generally made available.

In summary, this study presents a novel Bayesian network-based pipeline for predicting the genetic component of gene expression using GRNs reconstructed from the same data that are usually used for training transcriptome imputation models from cis-eQTL variants alone, providing important insights into the challenge of gene expression prediction for transcriptome-wide association studies.

\printbibliography

\newpage

\renewcommand\thesection{S\arabic{section}}
\renewcommand\thefigure{S\arabic{figure}}
\renewcommand\thetable{S\arabic{table}}
\renewcommand\theequation{S\arabic{equation}}
\setcounter{figure}{0}
 \setcounter{table}{0}
 \setcounter{section}{0}
\setcounter{equation}{0}

\begin{center}
    {\LARGE \textbf{Supplementary Information}}
\end{center}

\newpage

\section{Gene Expression Prediction Algorithm Using Gene Regulatory Networks} \label{main_algorithm}

\begin{algorithm*}[ht!]
\caption{Gene Expression Prediction using a Gene Regulatory Network. }\label{alg:gene_expression_prediction}
\begin{algorithmic}
\Require $G$ - Directed Acyclic Graph representing Gene Regulatory Network
\Require $E$ - Genotype-matrix, genotype of cis-eQTLs associated with each gene in $G$ from $N$ peoples (available in public database)
\Require $X$ - Expression value matrix for all genes in $G$, from $N$ same sample as in $E$
\Ensure $\Hat{X}$ - Empty dictionary to store predicted value of each gene in $G$
\Require $Prediction\_Model$ - A machine learning model to use for prediction.
\Require $UsePredictedExpression$ - A boolean flag to decide whether to use predicted expressions in subsequent predictions.

\Function{IsRootNode}{$A, H$}
    \State \Return \textbf{True} \textbf{if} $g_i$ has no incoming edges in $G$ \textbf{else} \textbf{False}
\EndFunction

\Function{PredictGeneExpression}{$G, E, X, UseParentGenotype, UseGrandParentGenotype, UsePredictedExpression$}
    \State Validate that $G$ is a DAG and handle errors
    \State $SortedNodes \gets$ Topologically sort nodes in $G$
    \State $\Hat{X}\gets \{\}$ Initialize predicted expressions
    
    \For{$i, g_i$ \textbf{in} \textbf{enumerate}($SortedNodes$)}
        \State $E_i \gets$ Genotype of eQTLs of gene $g_i$ from $E$
        \State $E_i^{p} \gets$ Genotypes of eQTLs of parent genes to $g_i$ from $E$
        \State $E_i^{gp} \gets$ Genotypes of eQTLs of both parents and grand parents  of $g_i$ from $E$
        \State $X_i^{p} \gets$ Expression of parents of $g_i$ from $X$ 
        \State $Input\_cis \gets$   $E_i$
        \State $\Hat{X}_i^{cis} \gets$ $Prediction\_Model(Input\_cis)$
        
        \If{\Call{IsRootNode}{$g_i, G$}}
            \State $\Hat{X}[i] \gets$ $\Hat{X}_i^{cis}$
        \Else
            \If{$UseParentGenotype$}
                \State $Input\_trans \gets$   $E_i^{p}$ 
            \ElsIf{$UseGrandParentGenotype$}
                \State $Input\_trans \gets$   $E_i^{pp}$ 
            \Else
                \State $Input\_trans \gets$  $X_i$
            \EndIf
        \EndIf
        $\Hat{X}_i^{trans} \gets$ $Prediction\_Model(Input\_trans)$
        \State $\Hat{X}[i] \gets$  $\Hat{X}_i^{cis}$ +  $\Hat{X}_i^{trans}$
    \EndFor

    \If{$UsePredictedExpression$}
        \State \Call{PredictGeneExpression}{$G, E, \Hat{X}, False, False,True$}
    \EndIf

    \State \Return $Prediction\_Model$
\EndFunction

\end{algorithmic}

\end{algorithm*}

\cleardoublepage

\section{Data processing details}

\subsection{DREAM5 data}

The DREAM5  Systems Genetics Challenge A consisted of 15 sub-datasets, each representing a different network. The sub-datasets were created through simulations involving different sample sizes (100, 300, and 999) across five distinct networks. Each network consisted of 1000 genes and their corresponding genotypes. In each sub-dataset, there was a one-to-one correspondence between each gene and a genotype variable, with 25\% of these genotype variables being cis-expression Quantitative Trait Loci (eQTL). For our analysis, we utilized the first 1000 gene network and dataset with 1000 samples.

To  identify the 250 cis-eQTLs we used the Kruskal-Wallis test to get pvalues and selected the top 25\% based on the strength of association between gene expression and genetic variants. 

The dataset is available at \href{https://www.synapse.org/}{https://www.synapse.org/}

\subsection{Geuvadis data}

We used the PLINK2 software with the "--export Av" flag to convert biallelic SNP genotypes to a matrix of (0/1/2/'NA')-values, representing the major homozygous genotype, the heterozygous genotype, the minor homozygous genotype, and missing data, respectively.


We removed any SNPs that had missing data in one or more samples. Additionally, we discarded SNPs where the minor allele frequency (MAF) was less than 5\% of the total allele frequency. In practice, multiple genes often share the same SNP as their most significant cis-eQTL. However, we also removed these genes due to warnings from the network reconstruction tool we are using. After removing SNPs that are the most significant eQTL for multiple genes, we obtained 2979 genes, each with one unique most significant cis-eQTL.


It is worth noting that although each gene is linked to at least one SNP, the majority of genes are associated with multiple SNPs. In order to reconstruct the gene regulatory network, we only took into account the most significant cis-eQTL for each gene (because our network reconstructiion tool requires one eQTL per gene). However, when training the gene prediction models, we can consider all eQTLs for each gene. 

\newpage

\section{Comparison of prediction methods}

\begin{figure}[H]
    \centering
    \textbf{A. DREAM}
    \vspace*{-2mm}
    \begin{center}
        \includegraphics[width=.45\linewidth]{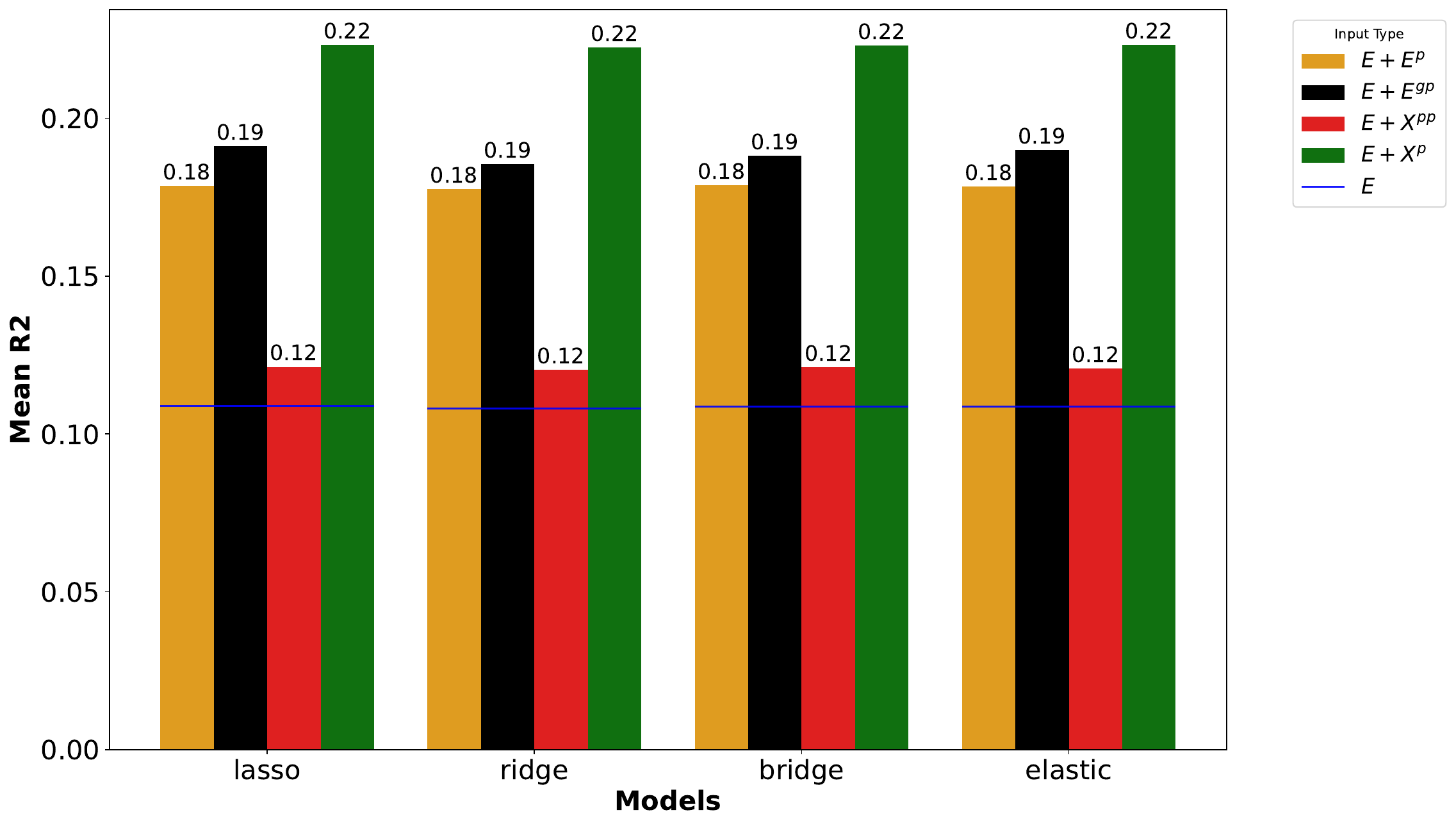}
    \end{center}
    
    \textbf{B. Yeast}
    \vspace*{-2mm}
    \begin{center}
        \includegraphics[width=.45\linewidth]{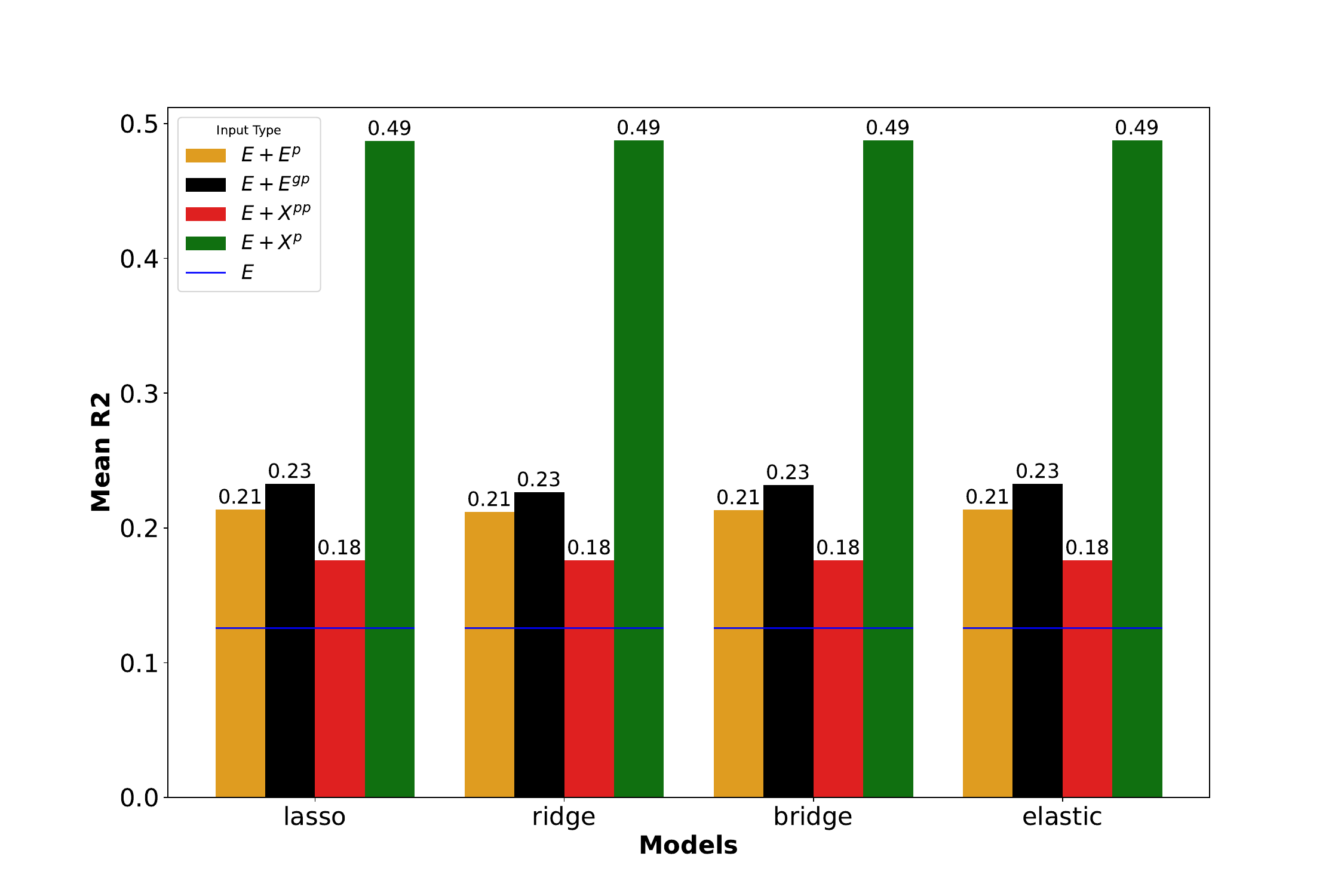}
    \end{center} 
    \textbf{C. Geuvadis}
    \vspace*{-2mm} 

      \begin{center}
        \includegraphics[width=.45\linewidth]{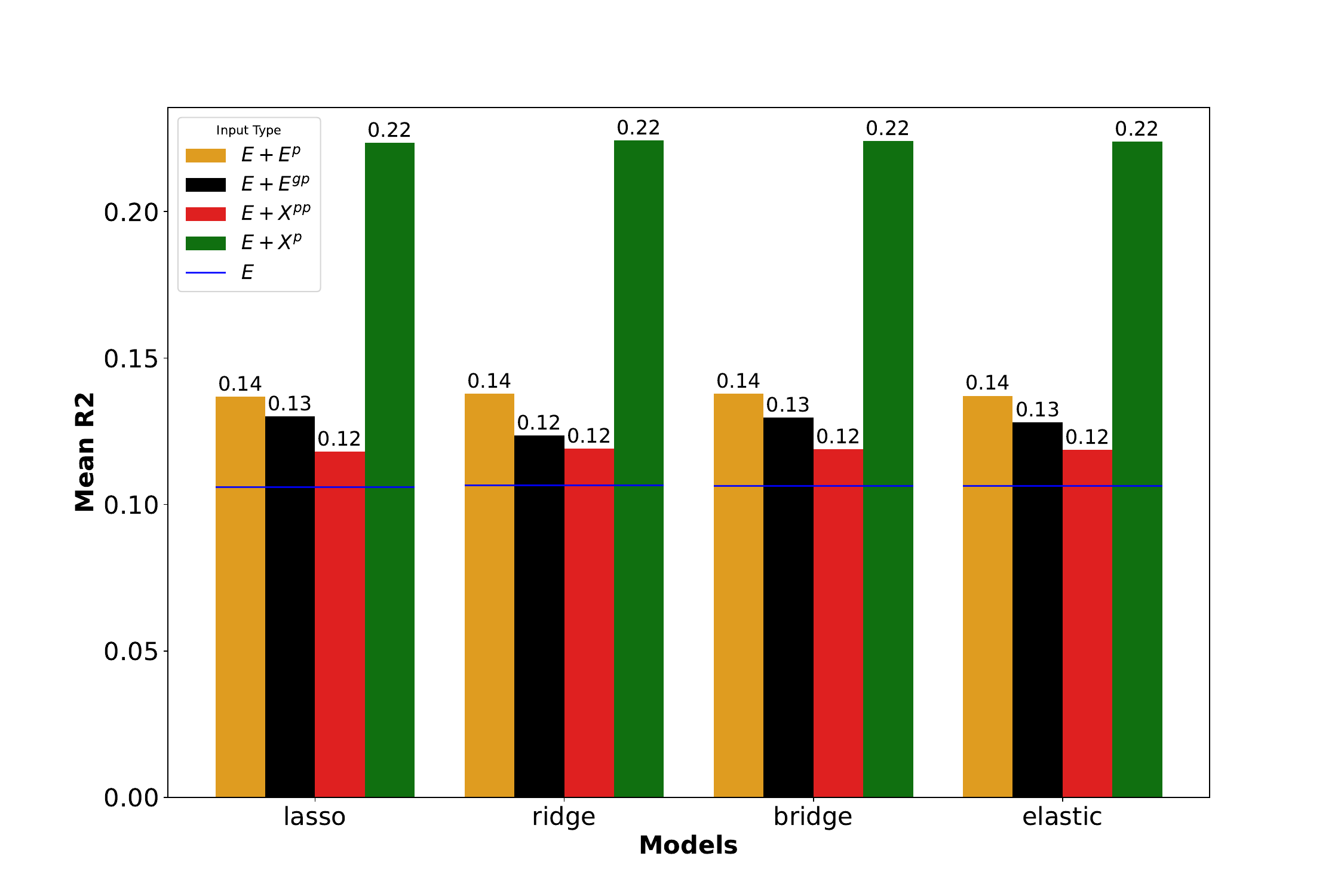}
    \end{center}

    \caption{\textbf{Comparison of Mean $R^2$ Values for Different Models Across DREAM, yeast, and Geuvadis Datasets}. The figures present the mean R² values (over GRN) for various regularized regression models predicting gene expression levels within Gene Regulatory Networks (GRNs) for three datasets: DREAM (A), yeast (B), and Geuvadis (C). Each figure displays the performance of Lasso, Ridge, Bridge, and Elastic Net models using different combinations of input features: cis-eQTLs alone (E), cis-eQTLs combined with parent eQTLs (E + Ep), parents and grandparents eQTLs (E + Egp), actual gene expression levels of parents (E + Xp), and recursively predicted parent expression levels (E + Xpp). }

    \label{fig:compare_model}
\end{figure}

\cleardoublepage

\section{Correlation Clustermap Analysis}\label{correlation_analysis}

\subsection{DREAM Dataset}
\begin{figure}[H]
    \centering
    \begin{center}
    \includegraphics[width = \linewidth]{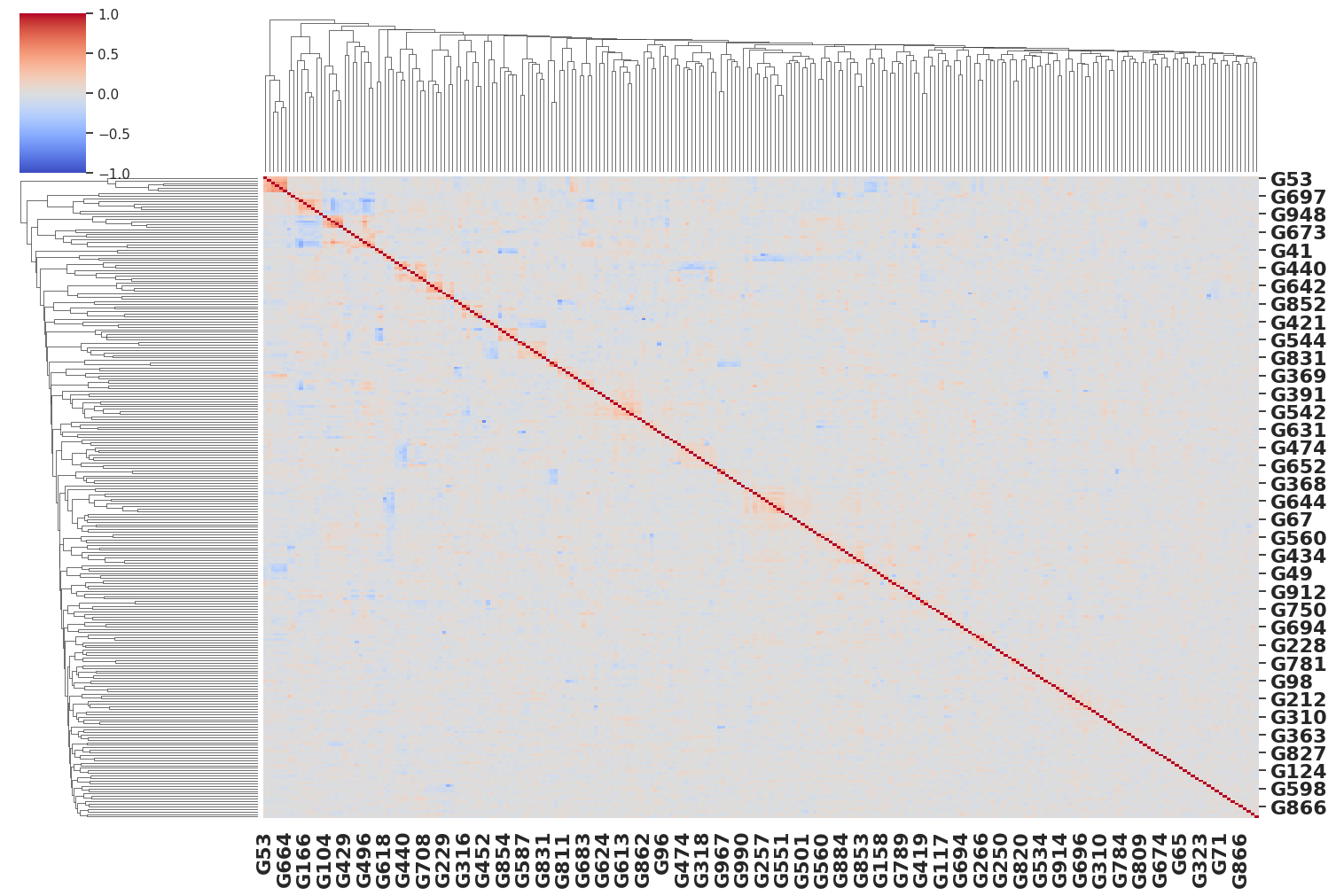}
    \end{center}
    \caption{\textbf{Gene Correlation Heatmap Clustermap for the DREAM Dataset}.
This heatmap illustrates the correlation between different genes in the DREAM dataset, with hierarchical clustering applied to group genes with similar correlation patterns. The predominantly weak red shading indicates a lack of strong gene interactions, and the clustering reveals a few less distinct clusters, suggesting a stochastic pattern of gene interactions..}
    \label{fig:corr_cluster_DREAM}
\end{figure}

Figure \ref{fig:corr_cluster_DREAM} presents the gene correlation heatmaps with hierarchical clustering for the DREAM dataset. The gene correlation heatmap for the DREAM dataset shows predominantly weak red shades, indicating weak positive correlations among genes. The hierarchical clustering suggests some grouping, but the clusters are not very distinct. This pattern suggests a more stochastic or random interaction among the genes, implying that the gene interactions in this dataset are not highly structured.

\subsection{Yeast Dataset}
\begin{figure}[H]
    \begin{center}
         \includegraphics[width = \linewidth]{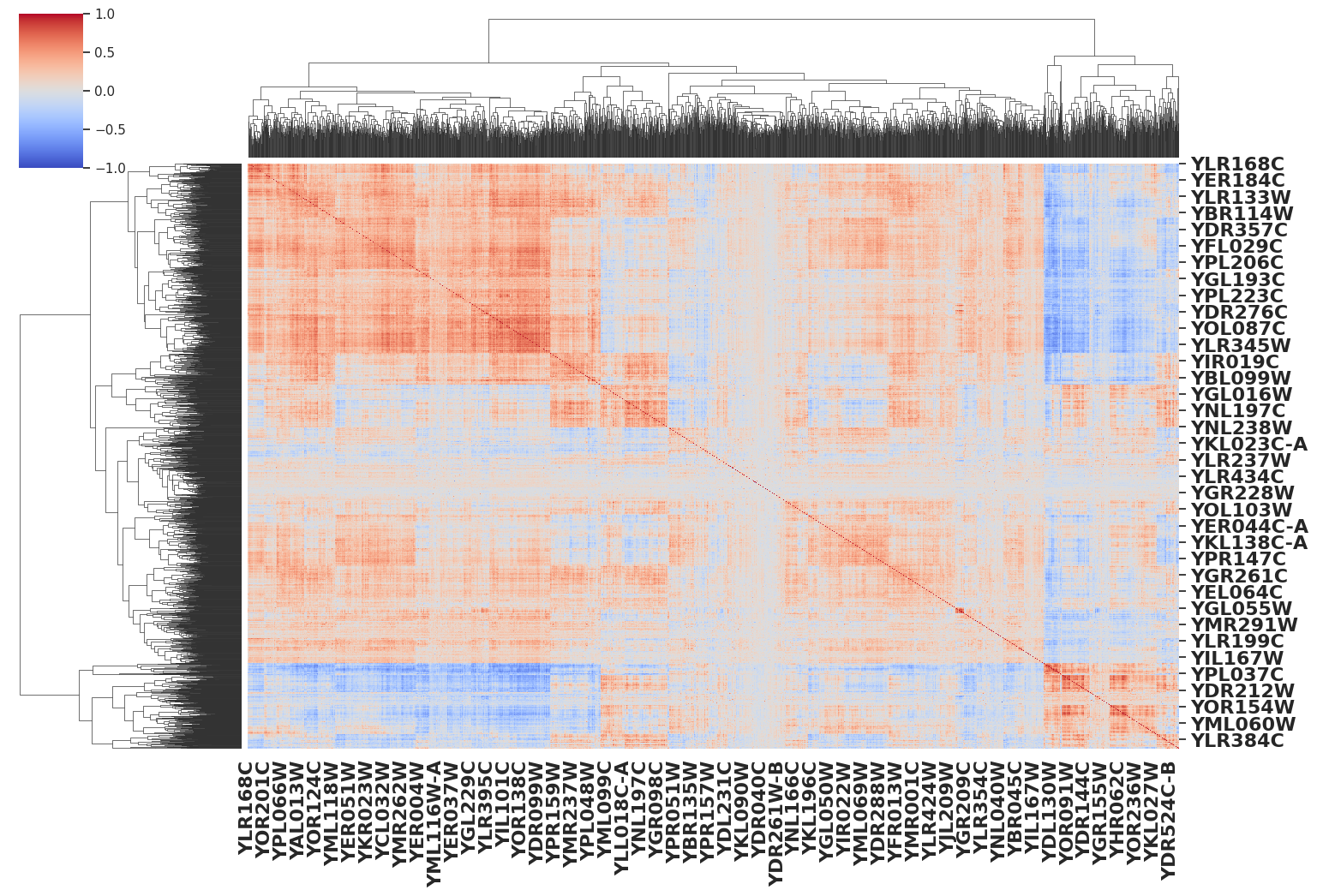}
    \end{center}
    \caption{ \textbf{Gene Correlation Heatmap Clustermap for the Yeast Dataset}.
This heatmap shows the correlation between genes in the yeast dataset, with hierarchical clustering highlighting groups of genes with similar correlation profiles. The clear patterns of positive (red) and negative (blue) correlations, along with well-defined clusters, indicate a high degree of structured gene interactions.}
        \label{fig:corr_cluster_yeast}
\end{figure}

Figure \ref{fig:corr_cluster_yeast} shows the gene correlation heatmaps with hierarchical clustering for yeast data. The gene correlation heatmap for the yeast dataset displays a clear pattern with strong positive (red) and negative (blue) correlations among genes. The hierarchical clustering reveals well-defined clusters, indicating groups of genes that have highly correlated expression profiles. This structured interaction implies a high degree of correlation organization within the yeast dataset, with certain genes interacting closely within specific clusters.

\subsection{Geuvadis Dataset}
\begin{figure}[H]
    \centering
    \begin{center}
         \includegraphics[width = \linewidth]{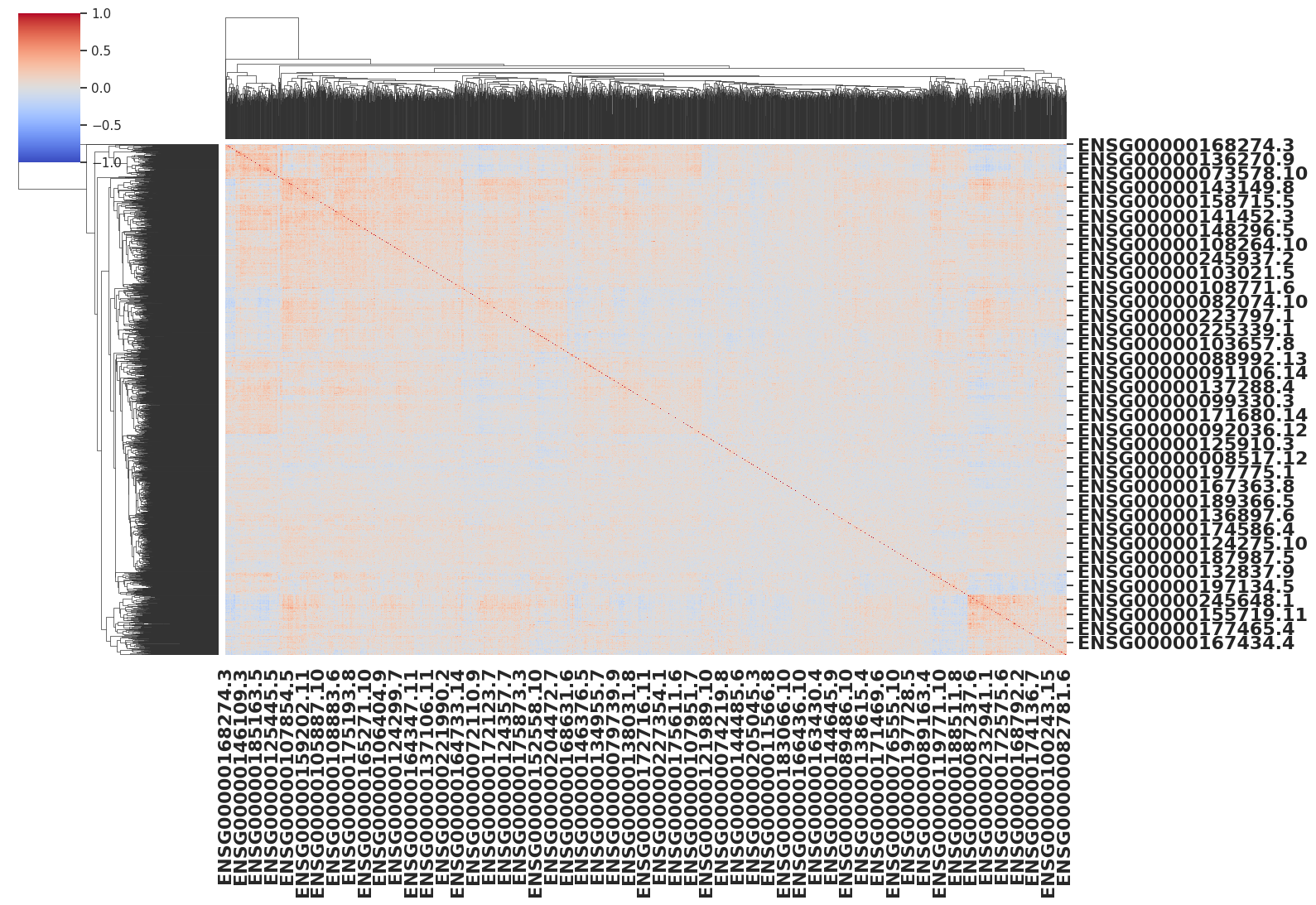}
    \end{center}
    \caption{ \textbf{Gene Correlation Heatmap Clustermap for the Geuvadis Dataset}.
This heatmap displays the correlation among genes in the Geuvadis dataset, with hierarchical clustering to identify clusters of genes with correlated expression. The predominantly weak red shading suggests weaker positive correlations and less structured gene interactions. The hierarchical clustering reveals clusters with more diffuse boundaries.}
        \label{fig:corr_cluster_geuv}
\end{figure}

Figure \ref{fig:corr_cluster_geuv} shows the gene correlation heatmaps with hierarchical clustering for Geuvadis data.The gene correlation heatmap for the Geuvadis dataset shows mostly weak red shades, indicating weak positive correlations. The hierarchical clustering reveals some clusters, but these clusters have more diffuse boundaries compared to the yeast dataset. This suggests that while there are some groups of correlated genes, the overall interaction structure is less pronounced than in yeast. The gene interactions in the Geuvadis dataset are less structured, indicating weaker regulatory relationships.

\section{Network Statistics}

\begin{table}[H]
    \centering
    \begin{subtable}{\linewidth}
        \caption{DREAM}
        \label{tab:network_comparasion_DREAM}
        \centering
        \begin{tabular}{l|ccc}
        \hline
         & $P$ & $P_2P_5$ & $P_0$ \\
        \hline
        edge posterior &  0.65 & 0.4 & 0.97\\
        global fdr     & 0.21 & 0.33 & 0.007 \\
        total  nodes    & 215 & 204& 211 \\
        num root nodes  & 41 & 43& 16 \\
        num leaf nodes & 50 & 50 & 70 \\
        num intermidate node & 124 & 111& 125 \\
        total edges      & 543& 447 & 706 \\
        \hline
        \end{tabular}
    \end{subtable}

    \vspace*{1mm} 
    \begin{subtable}{\linewidth}
    
        \caption{Yeast}
        \label{tab:network_comparison_yeast}
        \centering
        \begin{tabular}{l|rrr}
        \hline
       & $P$ & $P_2P_5$ & $P_0$ \\
       \hline
        edge posterior  &  0.9997 & 0.9994 & 0.99995  \\
        global fdr           & 0.0002 & 0.0004 & 0.000002  \\
        total  nodes          & 2107 & 2105 & 2399  \\
        num root nodes       & 495 & 489 & 337  \\
        num intermidate nodes & 1164 & 1152 & 1994  \\
        num leaf nodes       & 448 & 464 & 68 \\
        total edges          & 13888 & 13861 & 186008  \\
        \hline
        \end{tabular}
    \end{subtable} 

    \vspace*{1mm} 
    \begin{subtable}{\linewidth}
        \caption{Geuvadis}
        \centering
        \label{tab:network_comparison_geuv}
        \begin{tabular}{l|rrr}
        \hline
       & $P$ & $P_2P_5$ & $P_0$ \\
       \hline
        edge posterior &  0.7 & 0.5 & 0.9996  \\
        global fdr           & 0.23 & 0.36 & 0.0002  \\
        total  nodes          & 2495 & 2087 & 1887  \\
        num root nodes       & 81 & 93 & 474  \\
        num intermidate nodes & 684 & 472 & 1095  \\
        num leaf nodes       & 1730 & 1522 & 317 \\
        total edges          & 9018 & 5028 & 16020  \\
        \hline
        \end{tabular}
    \end{subtable} 
    
    \caption{Network Statistics Comparison for (a) DREAM5 data, (B) Yeast data and (c) Geuvadis Data}
    \label{tab:network_statistics_comparison}
\end{table}

\section{Using  All cis-eQTLs Geuvadis data}\label{snp_analysis}
\subsection{Comparison of prediction methods}

\begin{figure}[H]
    \centering
    \includegraphics[width=\linewidth]{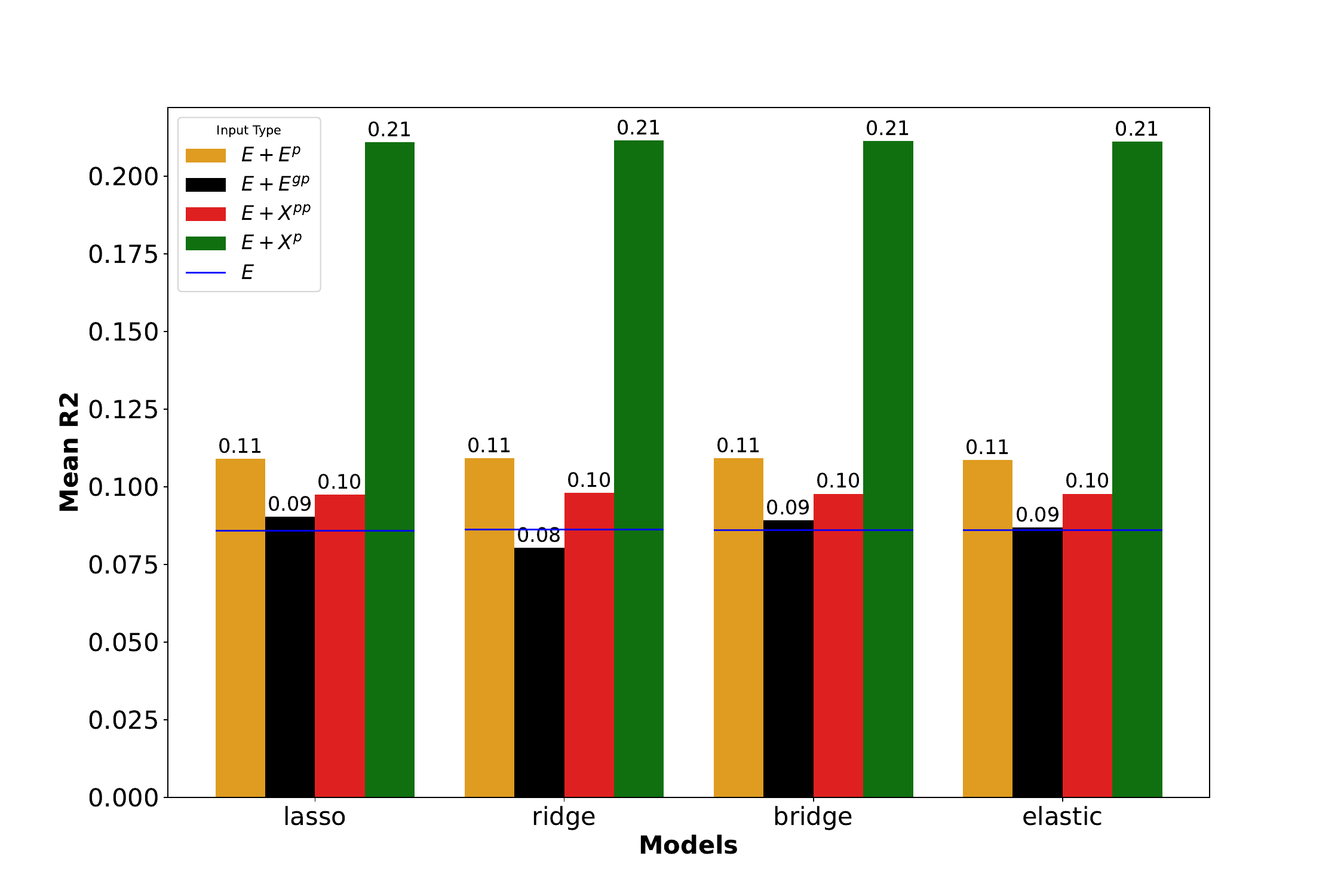}
    \caption{Comparison of Mean $R^2$ Values for Different Regularized Regression Models and Input Types on the Geuvadis Dataset Using All Cis-eQTLs}
    \label{fig:compare_model_appendix}
\end{figure}

Figure \ref{fig:compare_model_appendix} shows the performance of different regularized regression models (Lasso, Ridge, Bridge, and Elastic Net) on the Geuvadis dataset using various inputs, considering all cis-eQTLs rather than only the most significant eQTL. This approach aims to provide a more comprehensive understanding of the predictive power of all cis-eQTLs on low sample dataset. The mean $R^2$ values for each combination of model and input type are compared to understand the influence of different input types on prediction accuracy. 

The baseline model (E) have approximately $0.85$ across all models (Lasso, Ridge, Bridge, Elastic Net).

Mean $R^2$ of Parent eQTLs $(E + E^p)$ slightly better than the baseline, around $0,11$ for all models. The inclusion of parent eQTLs does n improve performance, indicating that for this dataset, parent eQTLs  provide  some additional predictive power when using all cis-eQTLs.

Mean $R^2$ of Grandparent eQTLs $(E + E^{gp})$, between $0.08 - 0.09 $ for all models. Meaning adding grandparent eQTLs do not enhance prediction accuracy in this dataset. Probabilily, due to overfitting due to large number of features.  

The results suggest that while  cis-eQTLs provide a solid foundation for gene expression prediction, the addition of parent and grandparent eQTLs with all cis eQTL does not contribute significantly in the when sample size is low.

\subsection{Relevance of network information} 

\begin{figure}[H]
    \centering
    \includegraphics[width=\linewidth]{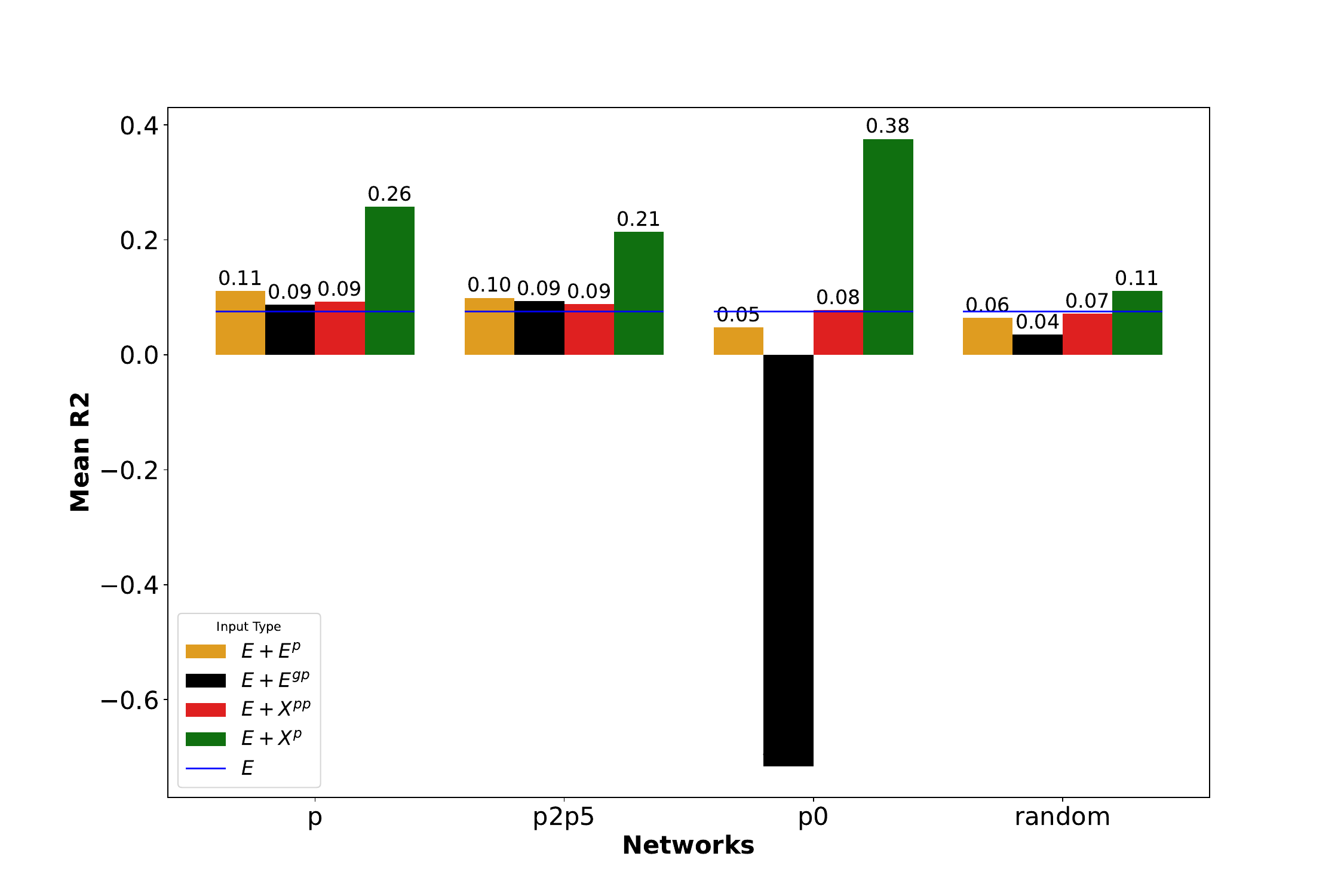}
    \caption{Comparison of Mean $R^2$ Values for Different Networks and Input Types on the Geuvadis Dataset Using All Cis-eQTLs}
    \label{fig:compare_net_appendix}
\end{figure}

Figure \ref{fig:compare_net_appendix} presents the performance of gene expression prediction models on the Geuvadis dataset using  different input types, where we use all eQTLs. The networks analyzed include the best causal network ($P$) and ($P_2P_5$), and a correlation network ($P_0$). The mean $R^2$ values are compared across these networks to understand the impact of different inputs and networks on prediction accuracy.

\begin{figure}[H]
    \centering
    \includegraphics[width=\linewidth]{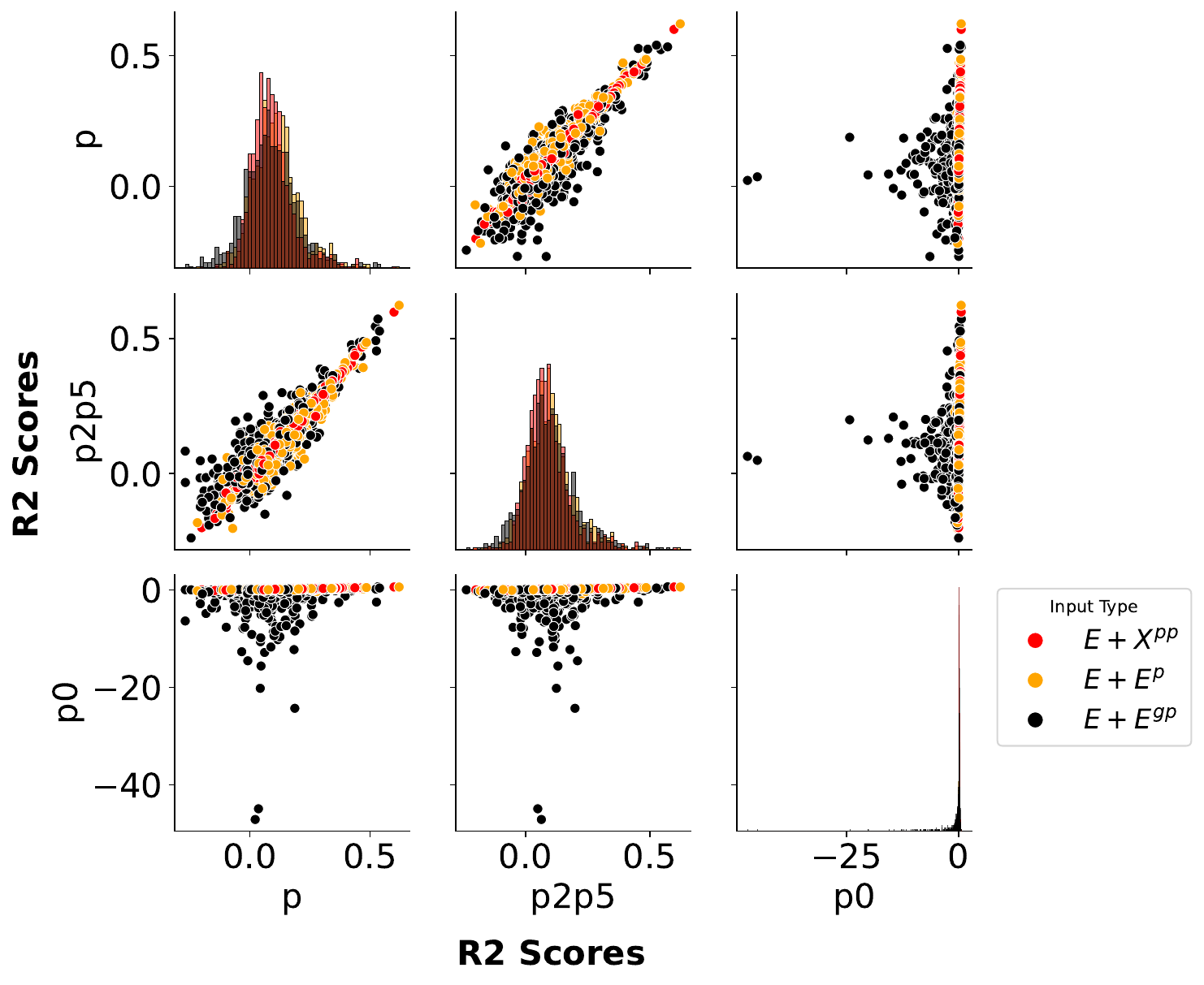}
    \caption{Pair Grid Comparison of $R^2$ Values for Different Networks and Input Types on the Geuvadis Dataset Using All Cis-eQTLs}
    \label{fig:compare_net_pair_appendix}
\end{figure}

Figure \ref{fig:compare_net_pair_appendix} shows the performance of different gene expression prediction models on the Geuvadis dataset using three different input types: cis-eQTLs (E), parent eQTLs $(E + E^p)$, and grandparent eQTLs $(E + E^{gp})$. The networks analyzed include the best causal network ($P$) and ($P_2P_5$), and a correlation network ($P_0$). The pair grid results provide a direct comparison of predictive performance for each pair of networks on the same genes and inputs by showing $R^2$ values for each predicted gene.

\begin{figure}
    \centering
    \textbf{A. DREAM}
    \vspace*{-2mm}
    \begin{center}
        \includegraphics[width=.55\linewidth]{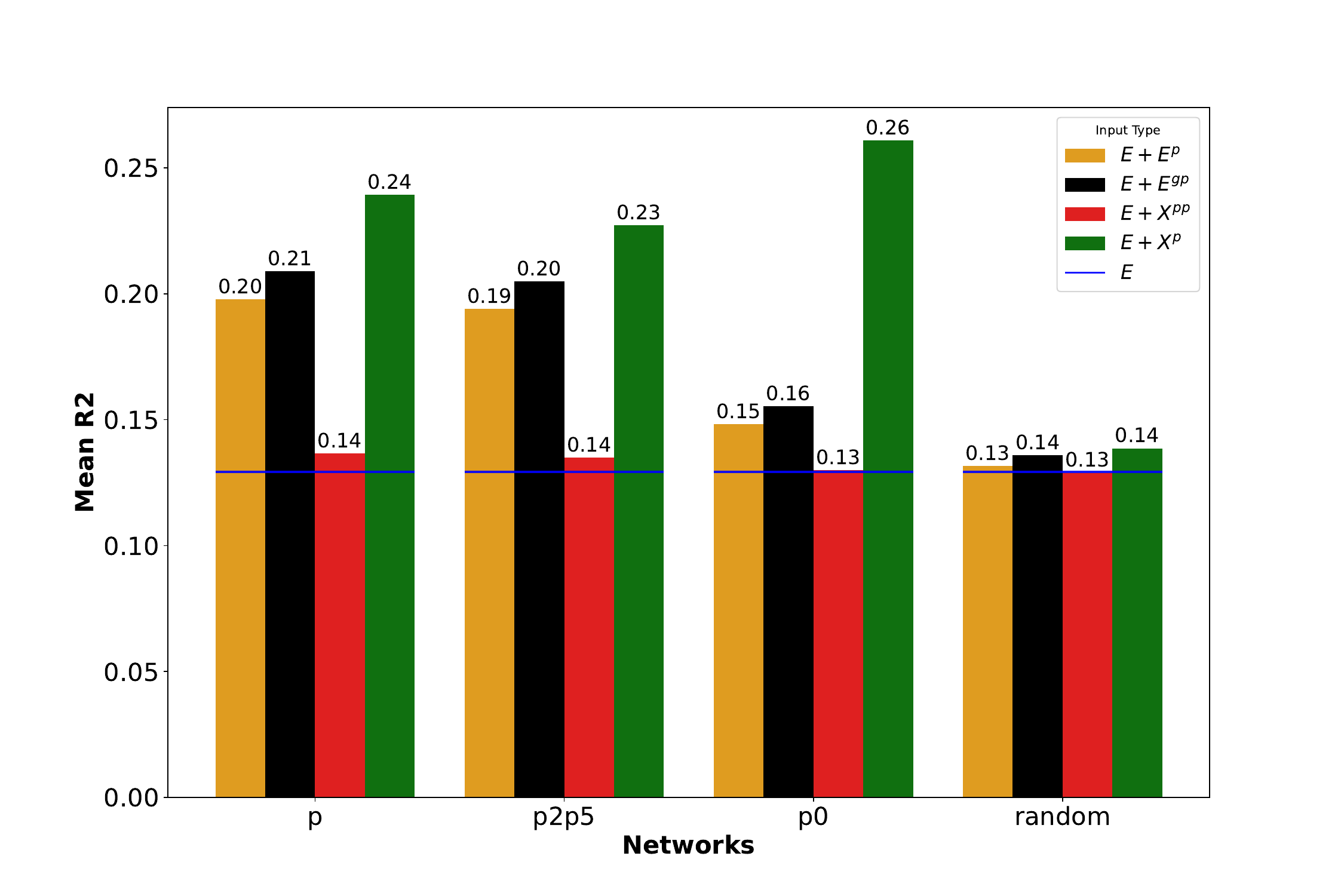}
    \end{center}
    \textbf{B. Yeast}
    \vspace*{-1mm}
    \begin{center}
        \includegraphics[width=.55\linewidth]{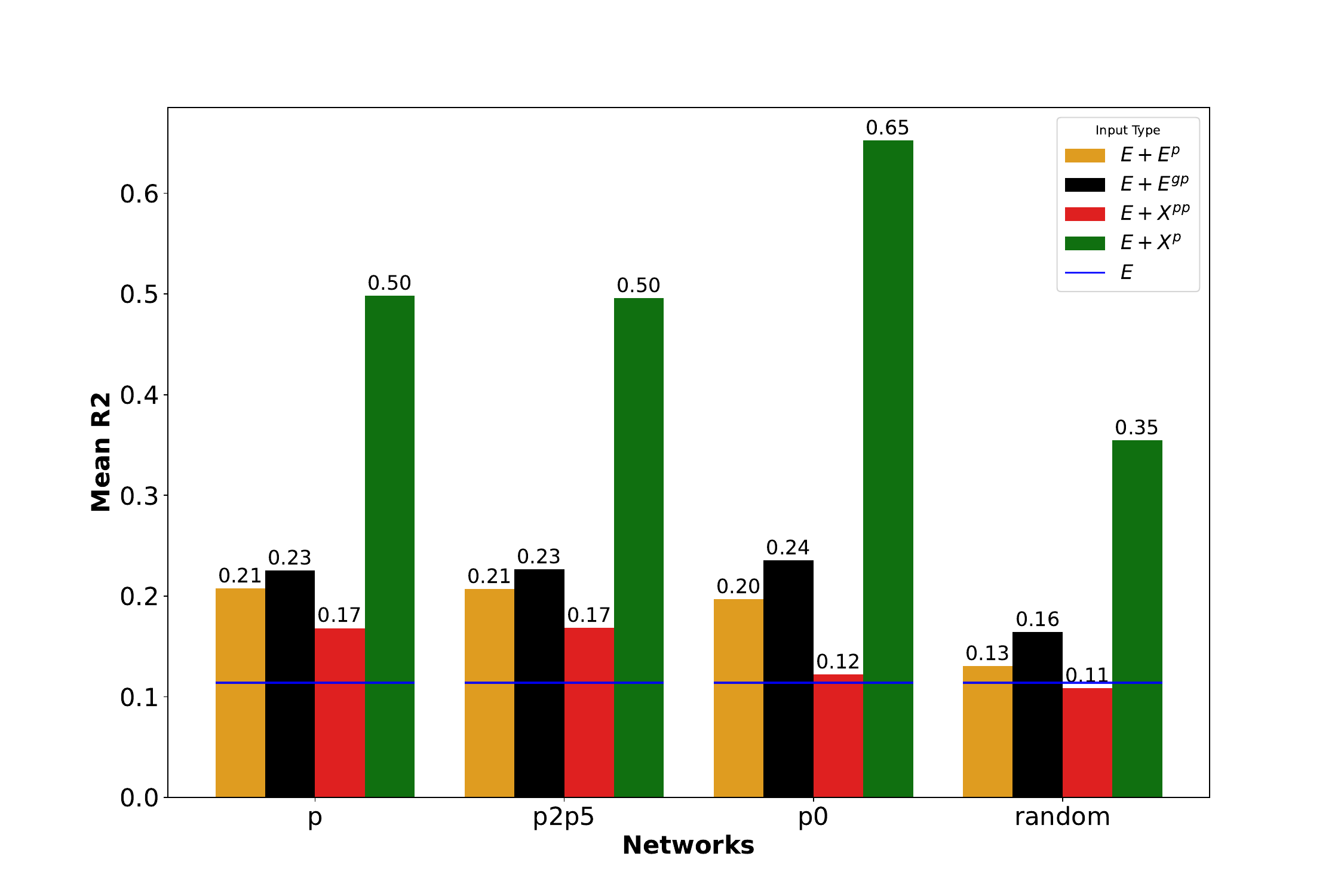}
    \end{center}

  \textbf{C. Geuvadis}
    \vspace*{-1mm}
    \begin{center}
        \includegraphics[width=.55\linewidth]{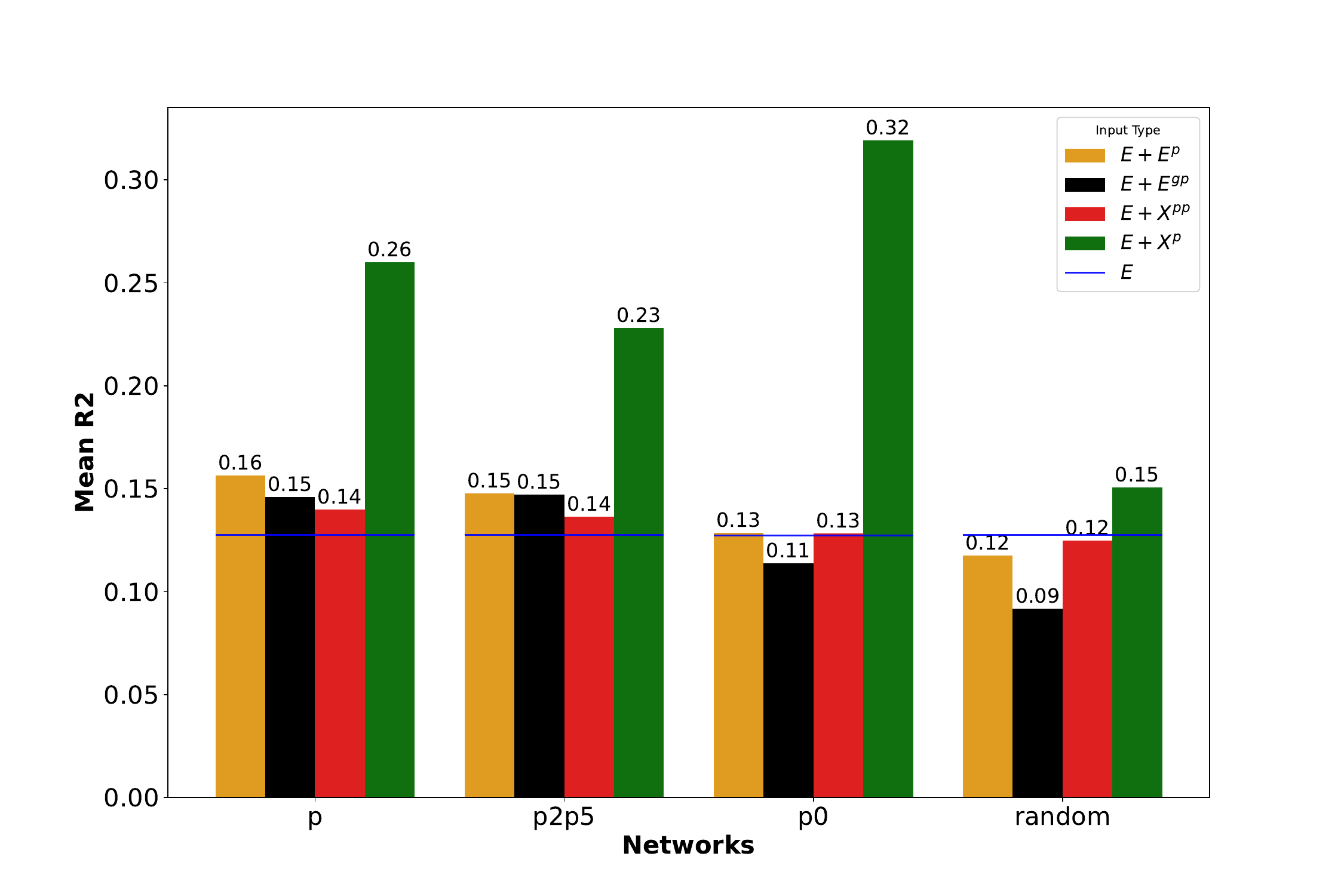}
    \end{center}
    
  \caption{\textbf{Assessment of the importance of network information.} Mean $R^2$ scores are shown of Bayesian Ridge regression using four reconstructed GRNs using Findr and five random networks (aggregated for plotting purposes), on DREAM data  (\textbf{A}) and Geuvadis (\textbf{B}) data.}
    \label{fig:compare_network_mean_bar}
\end{figure}

\end{document}